\documentclass[twocolumn]{aastex63}

\usepackage{graphicx}	
\usepackage{amsmath}	
\usepackage{amssymb}	

\received{}
\revised{}
\accepted{}
\submitjournal{ApJ}

\shorttitle{Jellyfish galaxy as astrophysical laboratory}
\shortauthors{Poggianti et al.}

\begin{document}






\title{GASP XXIII: A jellyfish galaxy as an astrophysical laboratory of the baryonic cycle}

\correspondingauthor{Bianca M. Poggianti}
\email{bianca.poggianti@inaf.it}

\author[0000-0001-8751-8360]{Bianca M. Poggianti}
\affiliation{INAF-Padova Astronomical Observatory, Vicolo dell'Osservatorio 5, I-35122 Padova, Italy}

\author[0000-0003-1581-0092]{Alessandro Ignesti}
\affiliation{Dipartimento di Fisica e Astronomia, Universita' di Bologna, via Gobetti 93/2, 40129 Bologna, Italy}
\altaffiliation{INAF, Istituto di Radioastronomia di Bologna, via Gobetti 101, 40129 Bologna, Italy}

\author[0000-0002-0843-3009]{Myriam Gitti}
\affiliation{Dipartimento di Fisica e Astronomia, Universita' di Bologna, via Gobetti 93/2, 40129 Bologna, Italy}
\altaffiliation{INAF, Istituto di Radioastronomia di Bologna, via Gobetti 101, 40129 Bologna, Italy}

\author[0000-0001-5840-9835]{Anna Wolter}
\affiliation{INAF-Brera Astronomical Observatory, via Brera 28, I-20121 Milano, Italy}

\author{Fabrizio Brighenti}
\affiliation{Dipartimento di Fisica e Astronomia, Universita' di Bologna, via Gobetti 93/2, 40129 Bologna, Italy} 

\author{Andrea Biviano}
\affiliation{INAF, Astronomical Observatory of Trieste, via Tiepolo 11, 34131 Trieste, Italy}

\author{Koshy George}
\affiliation{Faculty of Physics, Ludwig-Maximilians-Universit{\"a}t, Scheinerstr. 1, 81679, Munich, Germany}
\altaffiliation{Department of Physics, Christ University, Hosur Road, Bangalore 560029, India}
\altaffiliation{Indian Institute of Astrophysics, Koramangala II Block, Bangalore, India}

\author[0000-0003-0980-1499]{Benedetta Vulcani}
\affiliation{INAF-Padova Astronomical Observatory, Vicolo dell'Osservatorio 5, I-35122 Padova, Italy}

\author[0000-0002-7296-9780]{Marco Gullieuszik}
\affiliation{INAF-Padova Astronomical Observatory, Vicolo dell'Osservatorio 5, I-35122 Padova, Italy}

\author[0000-0002-1688-482X]{Alessia Moretti}
\affiliation{INAF-Padova Astronomical Observatory, Vicolo dell'Osservatorio 5, I-35122 Padova, Italy}

\author[0000-0001-9143-6026]{Rosita Paladino}
\affiliation{INAF-Istituto di Radioastronomia, via P. Gobetti 101, I-40129 Bologna, Italy}

\author[0000-0002-4158-6496]{Daniela Bettoni}
\affiliation{INAF-Padova Astronomical Observatory, Vicolo dell'Osservatorio 5, I-35122 Padova, Italy}

\author{Andrea Franchetto}
\affiliation{Dipartimento di Fisica e Astronomia, vicolo dell'Osservatorio 5, 35136 Padova, Italy}
\altaffiliation{INAF-Padova Astronomical Observatory, Vicolo dell'Osservatorio 5, I-35122 Padova, Italy}

\author[0000-0003-2150-1130]{Yara L. Jaff\'e}
\affiliation{Instituto de Fisica y Astronomia, Universidad de Valparaiso, Avda. Gran Bretana 1111 Valparaiso, Chile,}

\author[0000-0002-3585-866X]{Mario Radovich}
\affiliation{INAF-Padova Astronomical Observatory, Vicolo dell'Osservatorio 5, I-35122 Padova, Italy}

\author[0000-0003-2076-6065]{Elke Roediger}
\affiliation{E.A. Milne Centre for Astrophysics, Department of Physics and Mathematics, University of Hull, Hull, HU6 7RX, UK}

\author[0000-0002-8238-9210]{Neven Tomi\v{c}i\'{c}}
\affiliation{INAF-Padova Astronomical Observatory, Vicolo dell'Osservatorio 5, I-35122 Padova, Italy}

\author{Stephanie Tonnesen}
\affiliation{Flatiron Institute, CCA, 162 5th Avenue, New York, NY 10010, USA}

\author{Callum Bellhouse}
\affiliation{University of Birmingham School of Physics and Astronomy, Edgbaston, Birmingham, UK}

\author[0000-0002-7042-1965]{Jacopo Fritz}
\affiliation{Instituto de Radioastronomia y Astrofisica, UNAM, Campus Morelia, AP 3-72, CP 58089, Mexico}

\author{Alessandro Omizzolo}
\affiliation{Vatican Observatory, Vatican City, Vatican State}
\altaffiliation{INAF-Padova Astronomical Observatory, Vicolo dell'Osservatorio 5, I-35122 Padova, Italy}

\begin{abstract}
With MUSE, Chandra, VLA, ALMA and UVIT data from the GASP programme we study the multiphase baryonic components
in a jellyfish galaxy (JW100) with a stellar mass 
$3.2 \times 10^{11} M_{\odot}$ 
hosting an AGN.
We present its spectacular extraplanar tails 
of
ionized and molecular gas,
UV stellar light, X-ray and radio continuum emission.  This galaxy
represents an excellent laboratory to study the interplay between 
different gas phases and star formation, and 
the influence of 
gas stripping, gas heating, and 
AGN.  We analyze the physical
origin of the emission at different wavelengths in the tail, in
particular in-situ star formation (related to 
$\rm H\alpha$, CO
and UV emission), synchrotron emission from relativistic electrons
(producing the radio continuum)
and heating of the stripped
interstellar medium (ISM)
(responsible for the X-ray emission).  We
show the similarities and differences of the spatial distributions of
ionized gas, molecular gas and UV light, and argue that the
mismatch on small scales (1kpc) is due to 
different stages of the
star formation process.  We 
present the relation 
$\rm H\alpha$--X-ray surface brightness, which is steeper for
star-forming regions than for diffuse ionised gas regions with high
[OI]/$\rm H\alpha$ ratio.  We propose that ISM heating due 
to interaction with 
the intracluster medium 
(either for mixing,
thermal conduction or shocks) is responsible for the X-ray tail, 
the observed [OI]-excess and 
the lack of star formation in the
northern part of the tail.  We also report the tentative discovery in
the tail of the most distant (and 
among the brightest) currently
known ULX, a point-like ultraluminous X-ray source commonly 
originating in a 
binary stellar system powered
either by an intermediate-mass black hole or a 
magnetized neutron star.
\end{abstract}

\keywords{galaxies: evolution -- galaxies: clusters: general}




\section{Introduction}





Ram pressure stripping is considered the most efficient mechanism to remove gas from galaxies in galaxy clusters \citep{Boselli2006}. A multitude of observational studies have observed the smoking gun of this physical process at various wavelengths with different techniques, mostly H{\sc i}, $\rm H\alpha$ narrow band imaging, UV/blue light, and, more recently, integral field spectroscopy \citep{Kenney2004,Chung2007,Hester2010,Smith2010,Merluzzi2013,Yagi2010,Kenney2014,Fossati2016,Jachym2017,Consolandi2017,Gullieuszik2017,Moretti2018,Fossati2019,Bellhouse2019}. 

The most extreme examples of galaxies undergoing strong ram pressure are the so called "jellyfish galaxies" \citep{Smith2010,Fumagalli2014,Ebeling2014}.
They have extraplanar, unilateral debris visible in the
optical/UV light and striking tails of $\rm H\alpha$
ionized gas. Most of the $\rm H\alpha$ emission in the tails
is due to photoionization by massive stars born in situ in the
tail in dynamically quite cold $\rm H\alpha$-emitting clumps resembling giant and supergiant HII regions and complexes \citep[and 
references therein]{Poggianti2019}, with possibly some exceptions
(e.g. NGC4569, \citealt{Boselli2016}). 

Optical line-ratio diagnostic diagrams maps obtained with
integral-field spectroscopy show that ionization mechanisms other than
in-situ star formation (SF) are also at play in the tails, 
contributing especially to the interclump diffuse
emission \citep{Fossati2016,Poggianti2019}. Different optical line ratios depict a generally consistent picture, but provide significantly different values for the fraction of tail emission due to star formation or shocks/heating.

In \citet{Poggianti2019} we studied the optical ionization mechanisms in the tails of a significant 
sample of jellyfish galaxies (16 in total) from the GAs Stripping Phenomena in galaxies survey (GASP,\footnote{\url{http://web.oapd.inaf.it/gasp/index.html}} Poggianti et al. 2017b), finding that the tail
emission characteristics of the jellyfish galaxy JW100 are peculiar.
At odds with the majority of the other jellyfish
galaxies, star formation is not the obviously dominant ionization mechanism of the tail:  according to the
[OIII]5007/$\rm H\beta$ vs [OI]6300/$\rm
H\alpha$ diagnostic diagram, it has only a few star-forming clumps in the tail and large amounts of ionized gas with an [OI]6300 line excess. 

A high [OI]6300/$\rm H\alpha$ ratio is usually interpreted as a sign of the presence of
shocks \citep{Rich2011}, and shock-heated molecular hydrogen has been observed with {\it Spitzer} in some cluster galaxies undergoing ram pressure stripping \citep{Sivanandam2010, Sivanandam2014, Wong2014}.
Thermal heating of the
stripped gas where this meets the hot Intracluster Medium (ICM) is another possible source
of ionization, and its relevance might depend on the local ICM
conditions, which can be studied with X-ray observations. 
The exact source of the [OI] excitation in jellyfish tails
is currently unknown.
Understanding why JW100 is so special in its tail ionization mechanism
might be the key to understand under what conditions are stars forming
in the tails, and when they are not. 

The interaction with the hot X-ray emitting ICM is expected to be crucial to set the conditions of the gas in the tails. Such interaction might give rise to an X-ray tail \citep{Sun2010}, but so far there are only a few X-ray emitting ram pressure stripped tails observed. JW100 has deep {\sl Chandra} archive data, as well as a set of multiwavelength data obtained by the GASP project,
and therefore offers a great opportunity to study the relation between the properties of the stripped gas tail and those of the ICM.
The only other jellyfish for which a comparably rich multi-wavelength dataset is available is ESO137-001, a low mass (5-8 $\times 10^9 M_{\odot}$) galaxy in the Abell 3627 cluster with $\rm H\alpha$ and other optical emission lines, molecular gas and X-ray tails \citep{Sun2007,Sun2010,Sivanandam2010,Fumagalli2014,Fossati2016,Jachym2019}.
Apart from JW100 and ESO137-001, there are X-ray studies for ESO137-002 (also in Abell 3627, \cite{Sun2010}), and UGC6697 in Abell 1367 \citep{SunVikhlinin2005},
and weak/shorter tails reported in NGC4438, NGC4388 in Virgo and NGC4848 in Coma \citep{Sun2010}, plus an X-ray map of D100 in Coma shown in \cite{Jachym2017}. No strong X-ray tail has been detected in the Virgo cluster, and \cite{Sun2010} and \cite{Tonnesen2011} explained this evidence with the fact that the X-ray tail luminosity should increase with the ambient pressure, which is not very high in Virgo. 


The aim of this work is to shed some light on the physical mechanisms that create tails observable at different wavelengths in jellyfish galaxies, with
the ultimate goal of understanding when the process of ram pressure stripping in clusters can be observed in any given range of the electromagnetic spectrum. To do this, 
we perform a simultaneous analysis of the multi-wavelength dataset collected for JW100 by GASP.
In \S2 we introduce the galaxy JW100 and summarize its properties based on previous studies, describing its host cluster and its location within the cluster in \S2.0.1.
Section \S3 presents all the data used in this paper: MUSE, ALMA, VLA, $\it Chandra$ (\S3.1) and UVIT (\S3.2).
The results of a detailed analysis of the X-ray data are shown in \S4. 
In \S5 we discuss the spatial distribution and the physical origin of the emission at different wavelengths: optical emission lines, CO and UV in \S5.1, radio continuum in \S5.2 and X-ray in \S5.3. 
The X-ray point sources, namely the AGN and the ULX candidate, are discussed in \S6. Our results are summarized in \S7.

In this paper we use a \cite{Chabrier2003} IMF and the standard concordance 
cosmology parameters $H_0 = 70 \, \rm km \, s^{-1} \, Mpc^{-1}$, ${\Omega}_M=0.3$
and ${\Omega}_{\Lambda}=0.7$. At the JW100 cluster redshift (z=0.05509), this
yields 1$^{\prime\prime}$=1.071 kpc. The galaxy itself has a redshift z=0.06189.

\section{The galaxy}
JW100 (also known as IC5337) is an almost edge-on spiral galaxy in the cluster Abell
2626 (Fig.~1, Table~1).\footnote{Some works in the literature
  refer to JW100 as an S0 galaxy (e.g. \citealt{Wong2008}). Our analysis
  of the MUSE I-band light profile shows a Type II disk \citep{Freeman1970,Erwin2008} and favors
the hypothesis this is a (perhaps early) spiral (Franchetto et al. in prep.). Moreover, our stellar
population analysis with the SINOPSIS code \citep{Fritz2017}
shows that star formation was significant and
widespread throughout the disk until the ram pressure started to strip
the gas,
as it happens in spirals. However, given its inclination
(about 75 degrees) it is hard to assign a robust Hubble type.} 
Selected by \citet{Poggianti2016} as a stripping
candidate, it is one of the GASP jellyfish galaxies with the most striking ionized gas
tails and is the most massive galaxy of the GASP
sample with a stellar mass $3.2 \times  10^{11}$ $M_\odot$
\citep{Poggianti2017b}.\footnote{We note that the stellar mass was estimated in a slightly different manner in other papers, but values are consistent within the errors, i.e. \cite{Vulcani2018, Poggianti2019}.} The stellar and ionized gas kinematics obtained with MUSE were presented in \citet{Poggianti2017b} and a visual 3D representation can be seen at https://web.oapd.inaf.it/gasp/jw100.html.

JW100 hosts a central AGN (Seyfert2), that is detectable both in X-ray  \citep{Wong2008}
and from MUSE emission-line ratios (\citealt{Poggianti2017b}, see also ESO press release \#1725 https://www.eso.org/public/news/eso1725/). A detailed
MUSE analysis confirms that AGN photoionization models are required to explain its emission line
properties in the central region \citep{Radovich2019}. This work has also found a biconical
outflow extending for $\sim 2.5 \rm kpc$ in the North-West to
South-East direction with a velocity offset of
$\sim 250 \rm \, km \, s^{-1}$ and a bolometric AGN luminosity
estimated from the luminosity of the [OIII]5007 line of
$10^{43.9} \rm \, erg \, s^{-1}$ \citep{Radovich2019}.
The derived mass outflow rate  is low, $< 0.01$ $M_\odot$ yr$^{-1}$, in agreement with what is observed in AGN of similar luminosity.
In \S6 we will provide the AGN X-ray luminosity.

\begin{figure*}
\centerline{\includegraphics[width=4.5in,angle=-90]{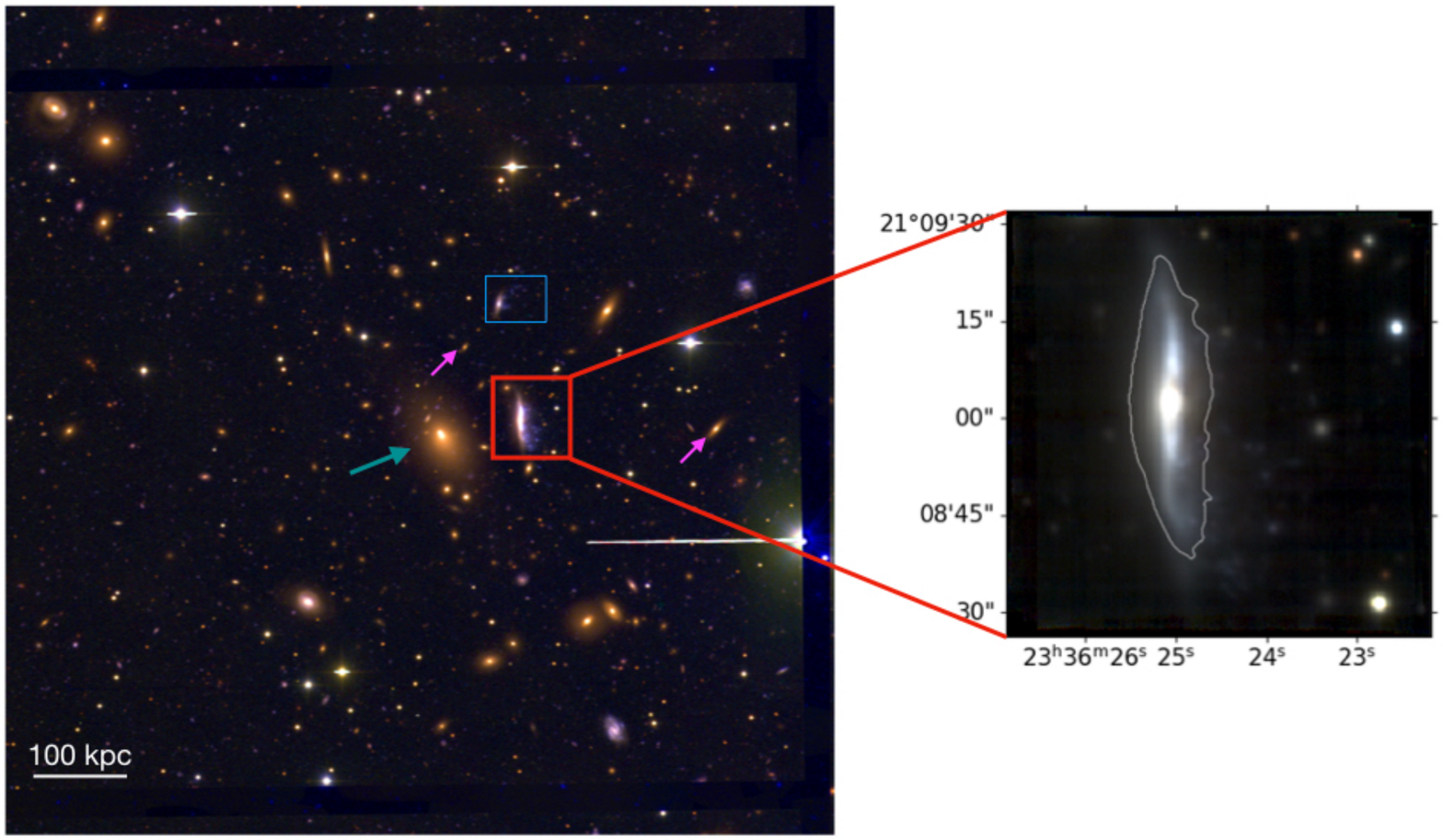}}
\caption{ RGB image of Abell 2626 (NUV-B-V, UVIT+WINGS) and a
  zoomed-in image of JW100 (gri MUSE). In the inset, the white contour represents 
the most external ($\sim 1.5 \sigma$ above background) 
isophote of the continuum MUSE light under $\rm H\alpha$
 and delineates the stellar disk. The major axis of this contour is 50
kpc.
The green arrow points to the BCG, the pink arrows point to the other two galaxies belonging to the JW100 substructure (\S2.0.1), and the blue rectangle identifies the other jellyfish, JW103.}
\end{figure*}

\citet{Poggianti2019} computed JW100's current star formation rate (SFR) from the $\rm H\alpha$
luminosity corrected both for stellar absorption and for dust 
extinction using the Balmer decrement adopting the \cite{Kennicutt1998a}'s relation 
($SFR (M_{\odot}/yr)=4.6 \times 10^{-42} L_{\rm H\alpha} (\rm erg/s)$) and including 
only those spaxels that according to the [OIII]5007/$\rm H\beta$ vs [SII]6717,6731/$\rm 
H\alpha$ diagram are ionized either by star formation.
We found a total (disk+tail) current star formation rate $SFR=4.0 M_{\odot}
yr^{-1}$, of which 20\% is in the tail. Its mass and SFR place JW100 
about 0.4dex below the SFR-mass relation for normal galaxies and 
$\sim 0.65$ dex below the relation for jellyfish galaxies \citep{Vulcani2018},
indicating that star formation has already decreased
due to gas stripping. 
When using the [OIII]5007/$\rm H\beta$ vs [OI]6300/$\rm H\alpha$ diagnostic diagram, due to the excess of [OI]6300 in the areas of diffuse emission in the tail (Fig.~2, see \S1), we find that the total SFR is only $SFR=2.0 M_{\odot}
yr^{-1}$, of which only 4\% is in the tail. The different conclusions reached using different optical emission lines will be discussed throughout the paper.

\begin{figure*}
\centerline{\includegraphics[width=3.5in]{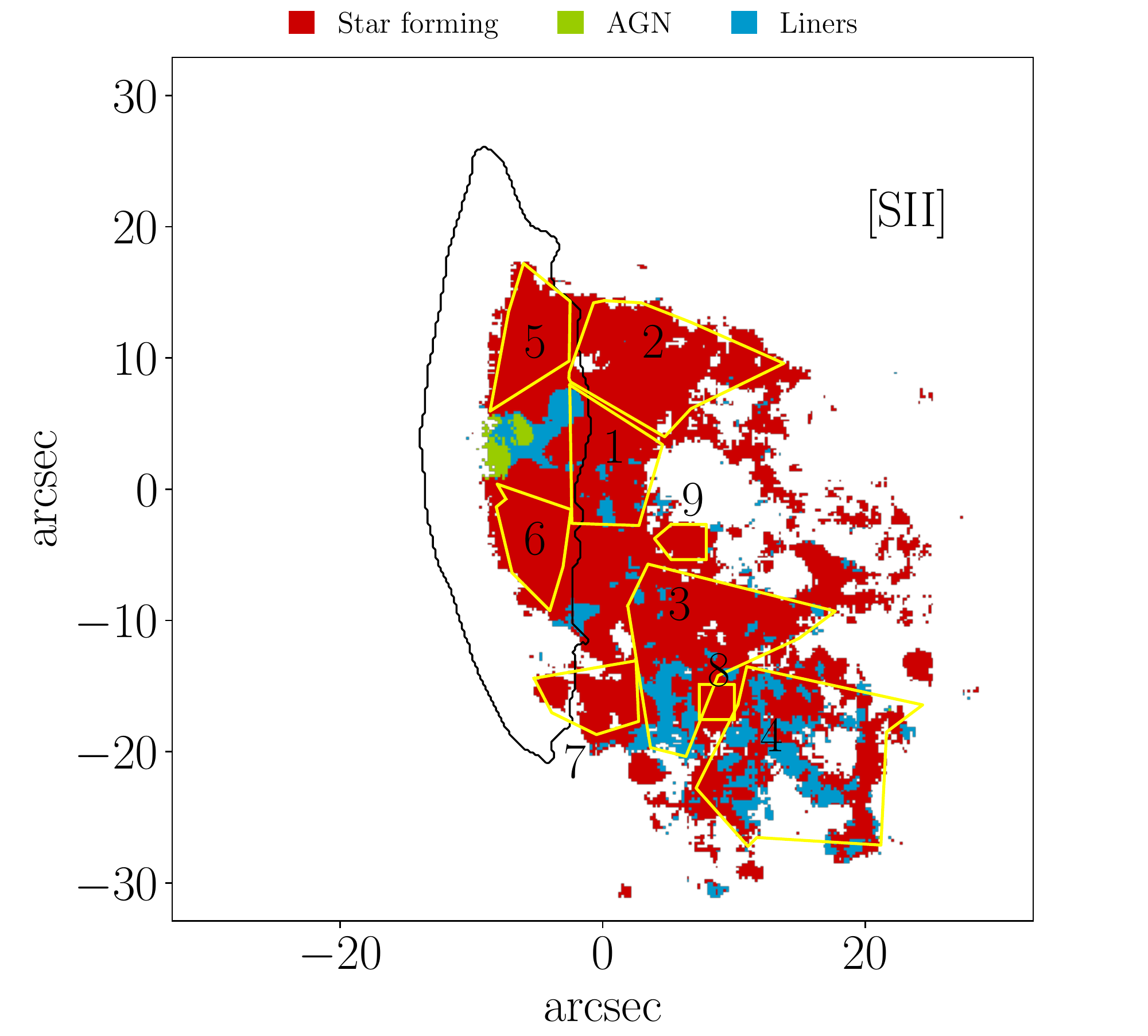}\includegraphics[width=3.5in]{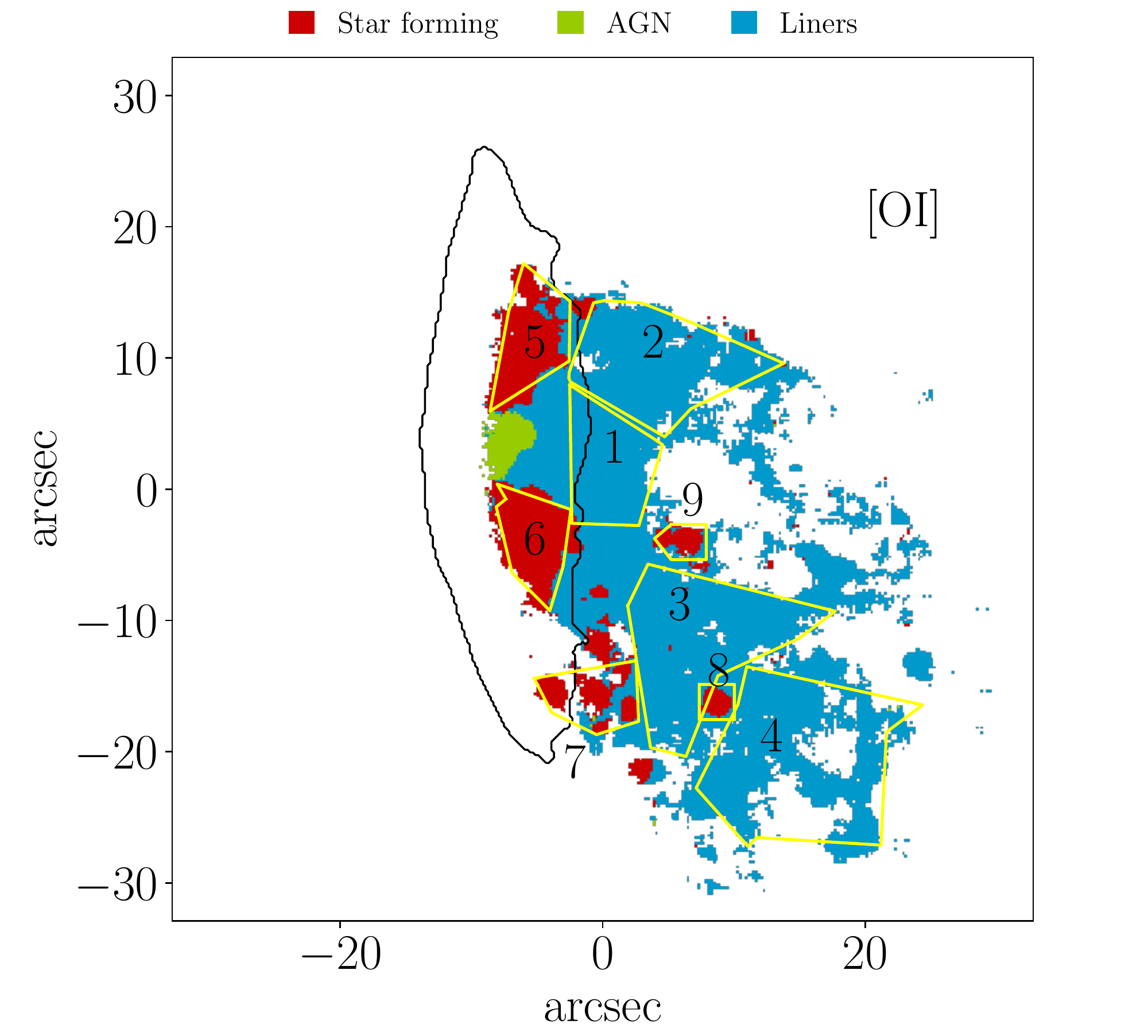}}
\caption{Ionization mechanism map according to (left) the [OIII]5007/$\rm
  H\beta$ vs [SII]6717,6731/$\rm H\alpha$ and (right) the [OIII]5007/$\rm
  H\beta$ vs [OI]6300/$\rm H\alpha$ diagnostic diagrams. Red=star
  formation. Green=AGN. Blue= LINER-like emission. The yellow poligons
define the regions that are studied in \S5.3. The black contour shows
the stellar disk as in Fig.~1.}
\end{figure*}

\subsubsection{JW100 in its cluster environment}
Abell 2626 
is a rather poor cluster with an estimated
X-ray luminosity of $1.9 \times 10^{44} \rm \, erg \, s^{-1}$ \citep{Wong2008}, a
velocity dispersion $\sigma=650^{+53}_{-49} \, \rm km \, s^{-1}$ and a mass $M_{200}=3.9_{-0.7}^{+1.5} \times 10^{14} M_{\odot}$ \citep{Biviano2017}. 
This cluster hosts a peculiar radio continuum emission with arc-like
features that appear quite symmetric around the cluster central cD
galaxy. The origin of this emission, also known as the ``Kite-radio source'', is still
unknown \citep{Gitti2004,Gitti2013,Ignesti2017,Kale2017,Ignesti2018}. 
JW100 is placed within A2626 in the most favourable conditions for ram
pressure stripping \citep{Jaffe2018}, with a very high line-of-sight velocity relative to the
cluster mean (1807 km s$^{-1}$ in the cluster rest-frame) and a projected
distance from the cluster center (taken to coincide with the Brightest
Cluster Galaxy, the cD galaxy IC5338) of only 83 kpc (Fig.~3). We note that $\sim 150$kpc to the North of JW100 there is another jellyfish candidate, JW103 \citep[and Fig.~1]{Poggianti2016}, and that the cD is also a very peculiar object, with a double nucleus and an AGN in the southern nucleus.

Using the OmegaWINGS spectroscopic catalogue of galaxies in the A2626 field \citep{Moretti2017}, we identify 92 members with a new membership algorithm (CLUMPS, Munari et al. in prep), based on the location of gaps in velocity space.
We then run the DS+ method of \citet{Biviano2017} on the selected
cluster members to detect cluster substructures. We detect 6
substructures that contain 21 cluster members in total. JW100 belongs
to one of these substructures, a group of three galaxies 
located at a
median distance of $100 \pm 62$ kpc from the cluster center, with a
median cluster rest-frame velocity of $1628 \pm 100$ km s$^{-1}$, and a
velocity dispersion\footnote{Being based on only
three values these statistical estimates must be considered very
tentative.} of $145_{-55}^{+79}$ km s$^{-1}$ (see Fig.~1). In Fig.~3 we display the projected phase-space
distribution of the cluster members within and outside substructures. 
The projected position and velocity of JW100 suggest that this galaxy is falling at very high speed into A2626 for the first time on a radial orbit in the direction opposite to the observer, and its likely close to pericentric passage \citep{Jaffe2018}. In addition, the extended tails visible in the plane of the sky indicate that the true velocity of the galaxy in the cluster must be higher than the (already high) measured line-of-sight velocity. 

Given the small projected distance between JW100 and the BCG, it is worth pondering the importance of
gravitational interactions between the two galaxies. First of all, it is worth noting that neither the deep optical MUSE image (see Fig.~1) nor the JW100 stellar velocity dispersion map (Fig.~1 in \cite{Poggianti2017b}) indicate a significant disturbance. The optical image shows a warped, regular disk and the stellar velocity map display a regular, undisturbed rotating disk. Second, the line-of-sight velocity of JW100 relative to that of the BCG is 1772$\, \rm km \, s^{-1}$ (from MUSE data of both), therefore this could only be a very high speed encounter. 
Moreover, our SINOPSIS spectrophotometric code does not detect a significant population of extraplanar old stars, that should be present if the extraplanar material were due to tidal effects. Furthermore, 
a crude approach to estimate the relative importance of the tidal acceleration $a_{tid}$ from a close neighbour versus the acceleration of the attraction from the galaxy itself $a_{gal}$ can be obtained, following \cite{Vollmer2005}, as $a_{tid}/a_{gal} = M_{BCG}/M_{JW100} (d/R -1)^{-2}$, where $R$ is the distance from the center of the galaxy and the BCG stellar mass $M_{BCG}= 7.8 \times 10^{11} \, M_{\odot}$ was estimated using literature absolute magnitude values and the \cite{BelldeJong2001} formulation. Assuming  as distance between the galaxies the projected distance $d \geq 83$kpc (which is a lower limit), the tidal acceleration is smaller than the gravitational acceleration from the galaxy itself out to 
$R=32$ kpc, larger than the stripping radius. Finally, the one-sided ionized gas tail and its direction with respect to the BCG disfavour the tidal hypothesis.
Although a mild tidal interaction cannot be excluded, we conclude that ram pressure stripping plays the major role for the points addressed in this paper.



\begin{figure}
\centerline{\includegraphics[width=3.5in]{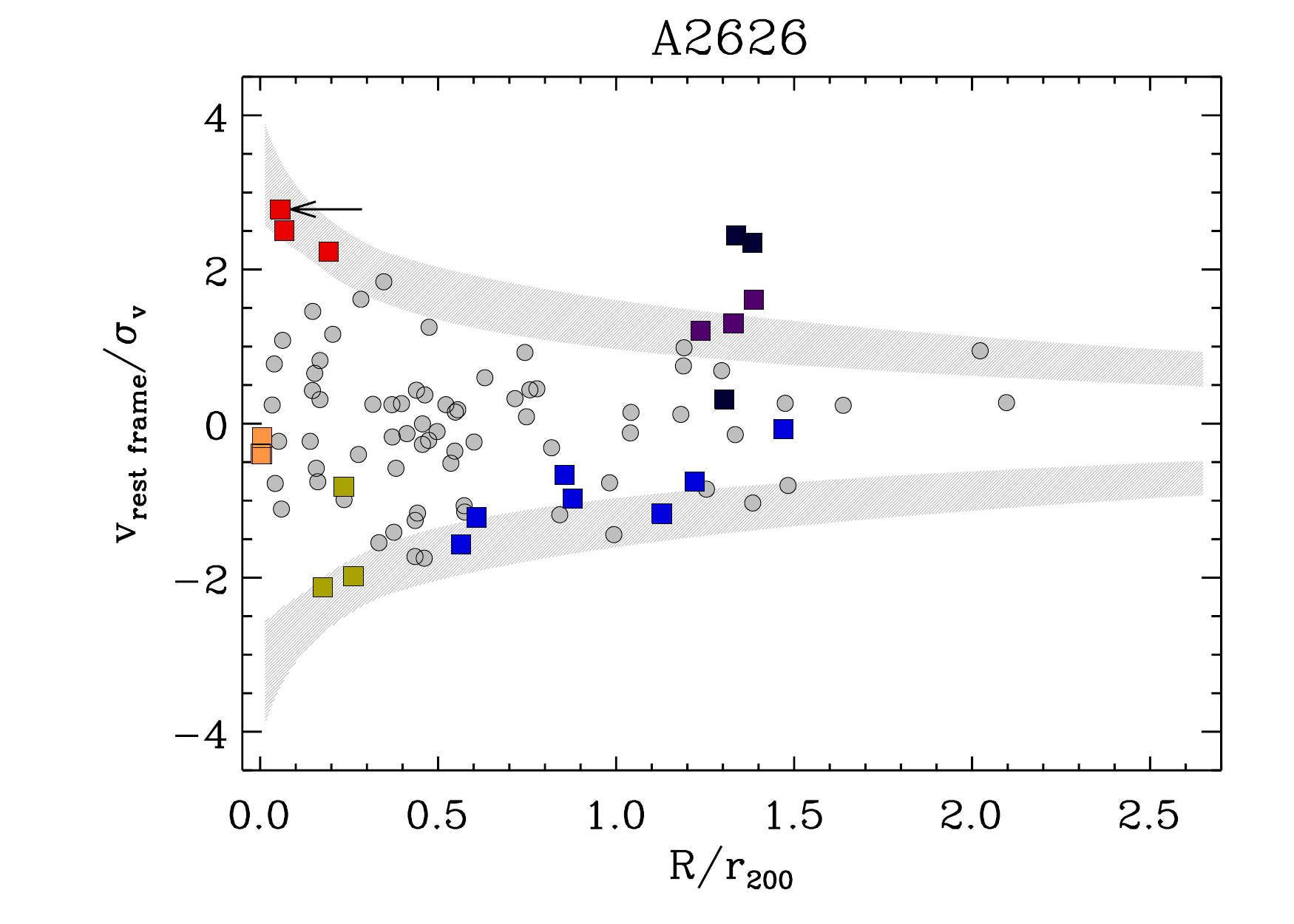}}
\caption{Distribution of cluster members in projected 
phase-space. Projected clustercentric distances $R$ and rest-frame line-of-sight velocities 
$v_{rf}$ are in units of the cluster virial radius and velocity 
dispersion, respectively (both taken from \citealt{Biviano2017}). 
Galaxies belonging to different substructures are represented by 
colored squares, each color defining a different substructure. The 
star symbol identifies JW100, which belongs to the substructure that includes also the two galaxies identified by the red squares. 
The grey-shaded regions represent the escape velocity curves with 1$\sigma$ uncertainties, assuming a \cite{Navarro1996} profile and \cite{Tiret2007} velocity anisotropy profile. The width of these regions take into account 1$\sigma$ uncertainties in the mass profile parameters $r_{200}$ and $r_{-2}$ from \cite{Biviano2017}, and allowing the \cite{Tiret2007} profile parameter $beta_{\infty}$ (see eq.8 in \cite{Biviano2013}) to vary from 0 (isotropic orbits) to 0.75 (radial orbits).
}
\end{figure}

\begin{table*}
\centering  
\caption{Columns are: 1) GASP ID number from \citet{Poggianti2016}; 
2) host cluster; 3) galaxy redshift; 4) cluster
 redshift from \citet{Moretti2014}; 5) cluster velocity dispersion from \citet{Biviano2017}; 6) and 7) galaxy RA and DEC; 8)
 galaxy stellar mass.
}
\begin{tabular}{lccccccc}
\hline  
$ID_{P16}$ & cluster & $z_{gal}$ & $z_{clu}$ & $\sigma_{clu} \, (km/s)$ & RA(J2000) & DEC(J2000)  
& $M_*(M_{\odot})$ \\
\hline  
JW100 	& A2626	 & 0.06189 & 0.05509 &   $650^{+53}_{-49}$ & 23:36:25.054 &	+21:09:02.64 &
$3.2^{+3.1}_{-1.2}  \times  10^{11}$  \\ 
\hline  
\end{tabular}
\end{table*}

\section{Data}

In this paper we use MUSE, ALMA, VLA, Chandra and UVIT data of JW100. 

This galaxy was observed as part of the GASP programme with the MUSE spectrograph in wide-field mode on July 15 2016 with $1\arcsec$ seeing, covering a 1'$\times$1' field-of-view with 0.2''$\times$0.2''
pixels and the spectral range between
4800 and 9300 \AA $\,$ sampled at 1.25 \AA/pixel with a spectral resolution
FWHM=2.6 \AA. \citep{Bacon2010}. The MUSE observations, data reduction and the methods of analysis are described in \citet{Poggianti2017}.
In the following, 
we will use the $\rm H\alpha$
emission (top left panel of Fig.~7) measured from the MUSE datacube corrected both for Galactic
foreground dust extinction and intrinsic dust extinction
using the measured Balmer decrement and for underlying stellar absorption using
our SINOPSIS stellar population code \citep{Fritz2017}. 
Moreover, we will use the ionization mechanism classification
based on the [OIII]5007/$\rm H\beta$ vs [SII]6717,6731/$\rm H\alpha$
and vs [OI]6300/$\rm H\alpha$ diagrams (Fig.~2) taken from \citet{Poggianti2019}.\footnote{The diagnostic diagram based on the [NII]6583/$\rm H\alpha$ ratio is not examined due to the contamination by a sky line.}

JW100 has been observed within the GASP project also with the Atacama Large mm/submm Array (ALMA) during Cycle 5.
Observations in Band 3 ($\sim$100 GHz) and band 6 ($\sim$220 GHz), were taken to sample the CO(1-0) and CO(2-1) lines, respectively, achieving in both bands a resolution of
$\sim$1 arcsec and an rms in 20 km/s wide channels of
$\sim$0.8 mJy/beam. The details of the ALMA observations, data reduction and analysis can be found in Moretti et al. (submitted).
In this paper we use the Band 6 data combined with 
additional Atacama Compact Array (ACA) observations,
sampling the angular scale from 1 to 26 arcsec.



We use the 
1.4 GHz A- and B-configuration VLA data from
a project focused on the peculiar kite radio
source located at the center of Abell 2626 (project code AG795, PI Gitti). These observations and the data reduction are described in \citet{Gitti2013}.
The VLA map at 1.4 GHz with a resolution of 3.8$\times$3.4 arcsec and an rms
of 15.6 $\mu$Jy beam$^{-1}$ is presented in the bottom right panel of Figure~7, where only regions at least 3$\sigma$ above 
the rms are displayed. 
By starting from the calibrated visibilities of \citet{Gitti2013},
we produced this image with CASA (Common Astronomy Software
application) v 4.7 by setting the 
visibility weighting to NATURAL and by adopting a tapering of the
baselines within 90 k$\lambda$ to enhance the sensitivity to the diffuse emission.

\begin{figure*}
\centerline{\includegraphics[width=6.7in]{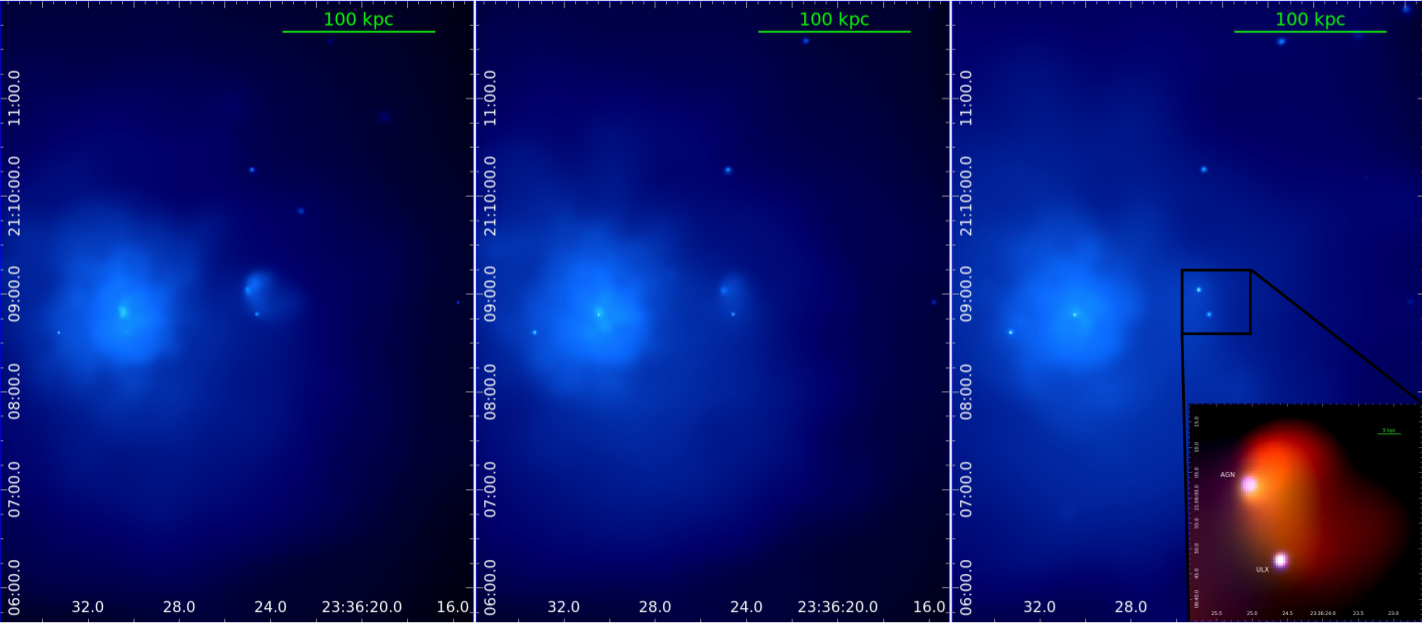}}
 \caption{Cluster-scale images showing the three X-ray bands (soft 0.5-1.2 keV (left), medium 1.2-2.0 keV (middle) and hard 2.0-7.0 keV (right)) of the field of JW100, smoothed with {\ttfamily csmooth} in CIAO. The inset in the rightmost panel is an RGB (red=0.5-1.2 keV, green=1.2-2.0 keV, blue=2.0-7.0 keV) smoothed zoom on the galaxy where the two point sources (AGN and ULX candidate) are clearly visible.
}
\end{figure*}

\subsection{\it{Chandra}} 
JW100 has been observed by the {\it Chandra} X-ray Observatory in two projects 
focused on Abell 2626 in January 2003 (ObsId: 3192, PI: C. Sarazin,  25 
ks exposure time) and in October 2013 (ObsID: 16136, PI: C. Sarazin, 110 
ks exposure time). The observations were made with the ACIS-S instrument 
in VFAINT mode. We retrieved the datasets from the {\it Chandra} 
archive\footnote{\url{ http://cxc.harvard.edu/cda/}} and reprocessed them 
with CIAO 4.10 and CALDB 4.8.1 to correct for known time-dependent gain 
and for charge transfer inefficiency.
In order to filter out 
strong background flares, we also applied screening of the event files.\footnote{\url{http://cxc.harvard.edu/ciao/guides/acis\_data.html}}

\begin{figure}
\centerline{\includegraphics[width=3.5in]{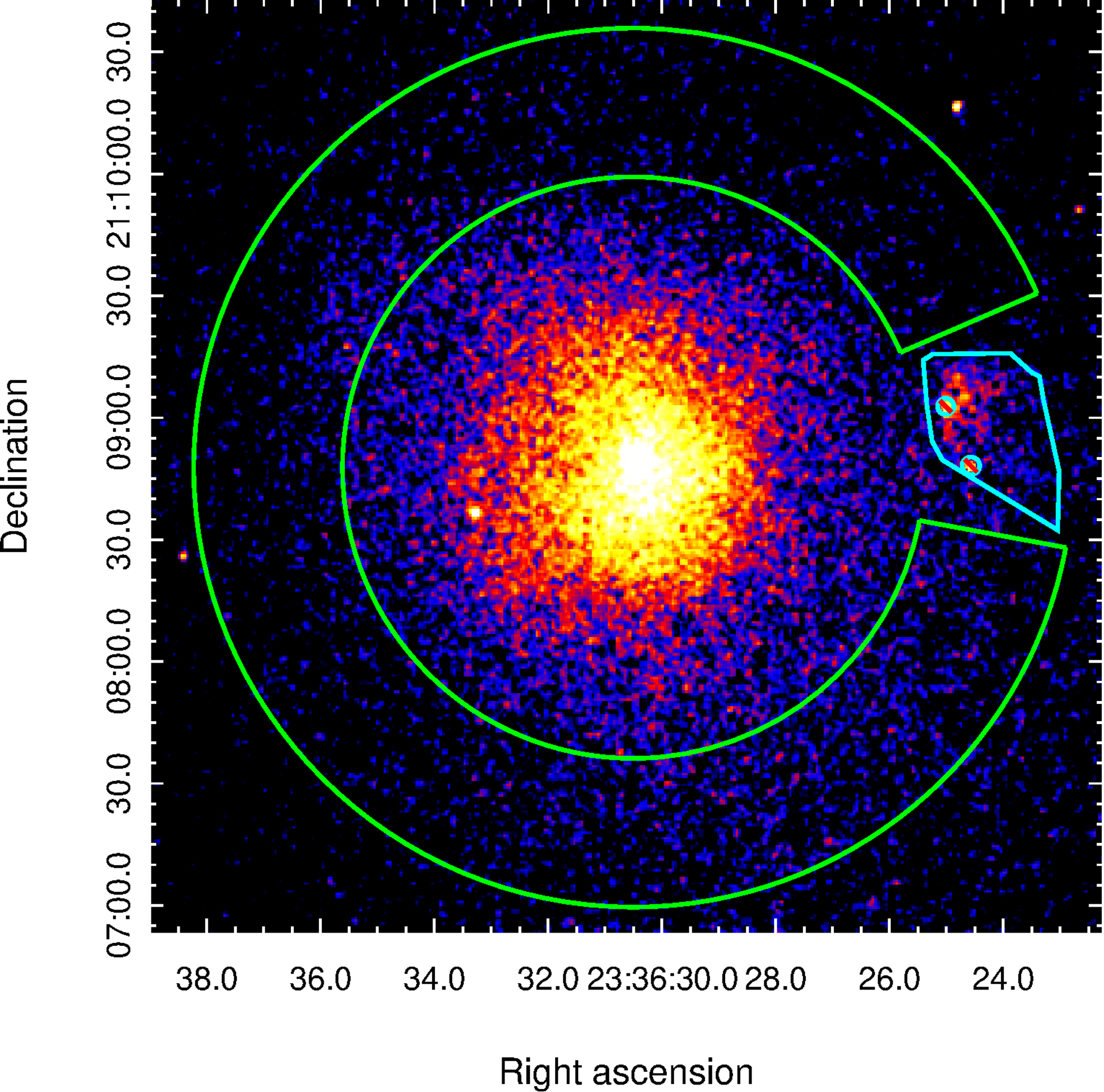}}
\caption{Background-subtracted, exposure-corrected Chandra image of A2626 in the 0.5-2.0 keV band smoothed with a 1.5" gaussian beam, with the galaxy (cyan) and control (green) regions highlighted. The two small circles within the JW100 region are the two point sources discussed extensively in \S6.
}
\end{figure}

For the background subtraction we used the CALDB "Blank-Sky" files 
normalized to the count rate of the source image in the 10-12 keV band. 
Finally, we identified the point sources using the CIAO task 
{\ttfamily WAVDETECT}, with the detection threshold set to the default 
value of $10^{-6}$ as probability to find a spurious source.
To improve absolute astrometry, we cross-matched 
the point sources identified in both datasets with the optical catalog 
USNO-A2.0, and then produced a mosaic of the two observations with the 
{\tt merge\_obs} script.

The exposure-corrected, background-subtracted Chandra 
mosaic in the 0.5-2.0 keV energy band with a resolution of $\sim$0.5
arcsec is shown in the bottom left panel of Fig.~7.

\subsection{UVIT}
The UV imaging of the GASP project is from the ultra-violet imaging telescope (UVIT)
onboard the Indian multi wavelength astronomy satellite ASTROSAT
\citep{Agrawal2006}. 
In this paper we use NUV imaging observations taken with the N242W
broad band filter, with an angular resolution of $\sim$ 1\farcs2.
\citep{Tandon2017a}. The NUV image is corrected for distortion
\citep{Girish2017}, flat field and satellite drift using the software
CCDLAB  \citep{Postma2017}. The final image created is for a net
integration of 10106.64s and is presented in the top right panel of Fig.~7.
The astrometric calibration is performed using the {\tt
 astrometry.net} package  where 
solutions are performed using USNO-B catalog \citep{Lang2010}. The
photometric calibration 
is done using the zero point values generated for photometric
calibration stars as described in  
\citet{Tandon2017b}. 
The UV data for JW100 and other GASP jellyfish galaxies will be used for a detailed analysis in a later paper.

\section{Results of the X-ray analysis}

While the analysis of the MUSE and ALMA data are presented elsewhere (Poggianti et al. 2017a, Poggianti et al. 2019, Moretti et al. submitted), in this section we describe the first detailed analysis of the X-ray data of JW100.

\subsection{Detection of point sources}
\label{sub:pointsrc}
We ran {\ttfamily WAVDETECT} on the two {\it Chandra} observations separately. 
There are only two point sources detected in the galaxy area,
whose position and source counts are listed in  Table~\ref{tab:pointsrc}. They are readily seen in excess to the galaxy diffuse emission in Figure~4,
in which the soft, medium and hard {\it Chandra} bands are shown. The contrast with the diffuse emission increases going from the soft to the hard band.  

The first source is positionally coincident with the center of the galaxy, and consistent with the AGN reported in \citet{Radovich2019}. The second source is just outside the optical extent of the stellar disc, but within the
X-ray tail. We remove these sources for the analysis of the X-ray diffuse emission described below.

\begin{table*} 
\centering  
\caption{Properties of the point sources. Positions, net counts, unabsorbed fluxes and luminosity. See \S~\ref{pointsrc} for details on the spectral shapes used.} 
\label{tab:pointsrc}
\begin{tabular}{lcccrrrrrr}
\hline  
Source & obsid &  RA(J2000) & DEC(J2000)  & net counts & net (0.3-2 keV) & net (2-10 keV) & f$_X ^{(0.5-2 keV)}$ & f$_X^{(2-10 keV)}$ &  L$_X^{(0.5-10 keV)}$\\
 & & & & & & & \multicolumn{2}{|c|}{ $\times 10^{-15}$ erg cm$^{-2}$ s$^{-1}$ }& $\times 10^{40}$ erg s$^{-1}$ \\
\hline  
AGN	& 3192 & 23:36:25.031 &	+21:09:02.53 & 43.8$\pm$7.3 & 22.9$\pm$5.3 & 21.7$\pm$4.8 & & & \\ 
AGN & 16136 &  & & 185.5$\pm$14.93 & 99.9$\pm$10.8 & 81.7$\pm$9.4 &2.2 & 32.7 & 23.8  \\
ULX? & 3192 & 23:36:24.592 & +21:08:47.65 & 35.8$\pm$6.7 & 28.9$\pm$5.8 & 9.7$\pm$3.3 & & & \\
ULX? & 16136 & & & 130.5$\pm$13.0 & 85.9$\pm$10.1 & 42.7$\pm$7.0 & 3.6 & 6.6 & 7.2\\
\hline  
\end{tabular}
\end{table*}

\subsection{Spectral analysis of the galaxy}
We performed a spectroscopic analysis of the {\it Chandra} data with XSPEC v 12.10 \citep{Arnaud1996}. 
We defined the region of interest of the spectral analysis, i.e. the galaxy, based on the MUSE observation to include the disk and stripped tail. Then, we defined a control region to study the properties of the ICM surrounding the galaxy. 
The ICM of Abell 2626 has an almost spherical symmetry \citep{Wong2008,Ignesti2018,Kadam2019},
thus, we expect all the thermal plasma at the same clustercentric distance of the galaxy to have similar properties\footnote{We neglect the increase in temperature due to the galactic Mach cone, because in projection it will be only a minor effect.}. To maximize the photons statistics, we used as control region a ring-shaped sector at the same distance of JW100. The galaxy region and the control region are shown in Fig. 5. 

We extracted a spectrum in each of the two regions using the CIAO task {\tt specextract} and then binned to give at least 25 counts in each energy bin. Similarly, we extracted the background spectrum from the Blank-Sky files in the same regions. The point sources 
were removed or masked (radius 1.8") during the spectrum extraction.

Spectra have been extracted separately from the two observations generating independent response matrices, and then, after  background subtraction, fitted jointly in the energy range 0.5-7 keV.


The control region spectrum was fitted with an absorbed thermal model ({\tt phabs*apec}) and the results are reported in Tab.~3. We measure a kT=$3.5\pm0.1$ keV , a metallicity Z=0.36$\pm0.04$ solar and an electron density $n_e=3.2\cdot10^{-3}$ cm$^{-3}$, that corresponds to a ICM density  $\rho_\text{ICM}$ of 5.8$\cdot 10^{-27}$ g cm$^{-3}$.
The properties of the ICM we derive here
are in agreement with previous results by \cite{Ignesti2018} and \cite{Kadam2019}.

As a first result, we ruled out that the ICM emission alone can reproduce
the observed emission from JW100 because a single-temperature model ({\tt apec} model) is not a good fit, as shown by the final statistics presented in Table~4 ($\chi^2=$173.43, DOF=96). 

Therefore, we modeled the spectrum extracted in the galaxy region as the combination of two components. To model
the cluster emission along the line of sight we used the absorbed, thermal, single-temperature component ({\tt apec}) described above
whose properties were fixed to that of the ICM measured in the control
region (Table~3). Then, to model the galactic emission itself, we adopted either another
single-temperature {\tt apec} model or a multi-phase, multi-temperature model, where the plasma emission measure  $EM=\int n_e n_\text{H} dV$, i.e. the normalization of the bolometric power emitted as thermal radiation, scales with the temperature as
$EM\propto T^{\alpha}$ and the temperature has an upper limit
$T_{max}$ (hereafter {\tt cemekl} model, Singh 1996).

The former is a simplified model where the galactic medium is a plasma emitting at a single temperature which is different from that of the local ICM.
The latter model is appropriate for a scenario in which the galactic X-ray emission comes from a multi-temperature plasma, that could be produced by the mixing of the hot ICM and the cold ISM triggered by the ISM stripping, the thermal conduction heating or the shock heating. In this case,
we may expect the temperature of the emitting
plasma to range from the temperature of the ICM to the temperature of
the ISM.
The photon statistics was not sufficient to obtain a solid estimate of the metallicity
of the plasma, so we fixed it at the solar value, which is  the metallicity of the stripped gas measured from the MUSE data.

We report the
results of the fits in Table~4.  With the double {\tt apec} model
($\chi^2=$93.58, DOF=95) we recover a temperature of
0.82$_{-0.05}^{+0.16}$ keV for the galactic component, which is lower
than the ICM.
In the
{\tt apec+cemekl} model we fitted the data at first by setting the
{\tt T$_{max}$} parameter to match the temperature of the ICM ($\chi^2=$93.84, DOF=95), then by letting it free ($\chi^2=$87.18, DOF=94). In the second case we recovered an
upper limit of the temperature {\tt T$_{max}$}=1.2$_{-0.26}^{+0.50}$
keV, which is lower than the ICM temperature.
The two models (double {\tt apec} vs {\tt apec+cemekl}) are statistically indistinguishable and they fit equally well the observations.

For each model we measured the unabsorbed X-ray luminosity in the 0.5-2.0, 0.5-10.0 and 0.3-10 keV bands associated to the galactic spectral component, listed in Table~4.

These findings will be discussed in \S5.3. 

 
\subsection{Search for the bow shock}
 
JW100 has a line-of-sight velocity of 1807 km s$^{-1}$ with respect to the cluster  and, based on the orientation of the H$\alpha$ tail we expect the total velocity to have also a significant trasversal component. From the values of the thermal properties of the ICM measured in the control region  we estimate a local sound velocity  $c_\text{s}\simeq1.5\cdot10^{4}T_\text{ICM}^{1/2}\simeq960$ km s $^{-1}$. 
Therefore, the galaxy is moving supersonically (with a tentative lower limit for the galaxy Mach number $\cal{M}\simeq $2) and, thus, we may expect to observe two discontinuities in front of it, the leading edge of the shock, i.e. the bow shock, and the contact discontinuity that drives this shock. Measuring the
jump temperature across the shock front could give us an independent measure of the galaxy Mach
number, thus of its velocity with respect to the ICM. We note that bow shocks in front of jellyfish cluster galaxies have never been observed
(but see \citet{Rasmussen2006} for the temperature jump in NGC 2276 in a galaxy group).

 We performed a morphological analysis to search for a brightness discontinuity in front of the infalling galaxy by adopting several geometries, finding
indications of a surface brightness jump at $\sim 6 " \sim 6$ kpc from the galaxy with a significance of 2$\sigma$
(Fig.~6). To have a spectroscopic confirmation, we further measured the temperature profile across the surface brightness jump. We
extracted the spectra in the 0.5-7.0 keV band in the supposedly post-shock (orange) and pre-shock (blue) regions across the brightness edge (Fig. 6) and we collected 1700 and 650 photons in the
outer and inner sectors, respectively. 
Our spectral
results
 may suggest a temperature jump at the
shock front ($kT_\text{pre}$ = 4.33$^{+0.30}_{−0.20}$ keV, $kT_\text{post}$ = 4.88$_{-0.39}^{+0.56}$ keV), although given the uncertainties the
pre-shock and post-shock regions are still consistent with being
isothermal. From this analysis we therefore conclude that the {\it Chandra} data
can neither confirm nor deny the existence of the shock front. 
This may be caused by the combination of the low data 
statistics and the complex morphology of the shock, as suggested by 
the H$\alpha$ surface brightness distribution. We note that with the expected mach number $\cal{M}\gtrsim$2 we would have a physical temperature jump around 2. However, projection effects would significantly decrease the jump and smear out the discontinuity.

\begin{figure*}
\centerline{\includegraphics[width=7.0in]{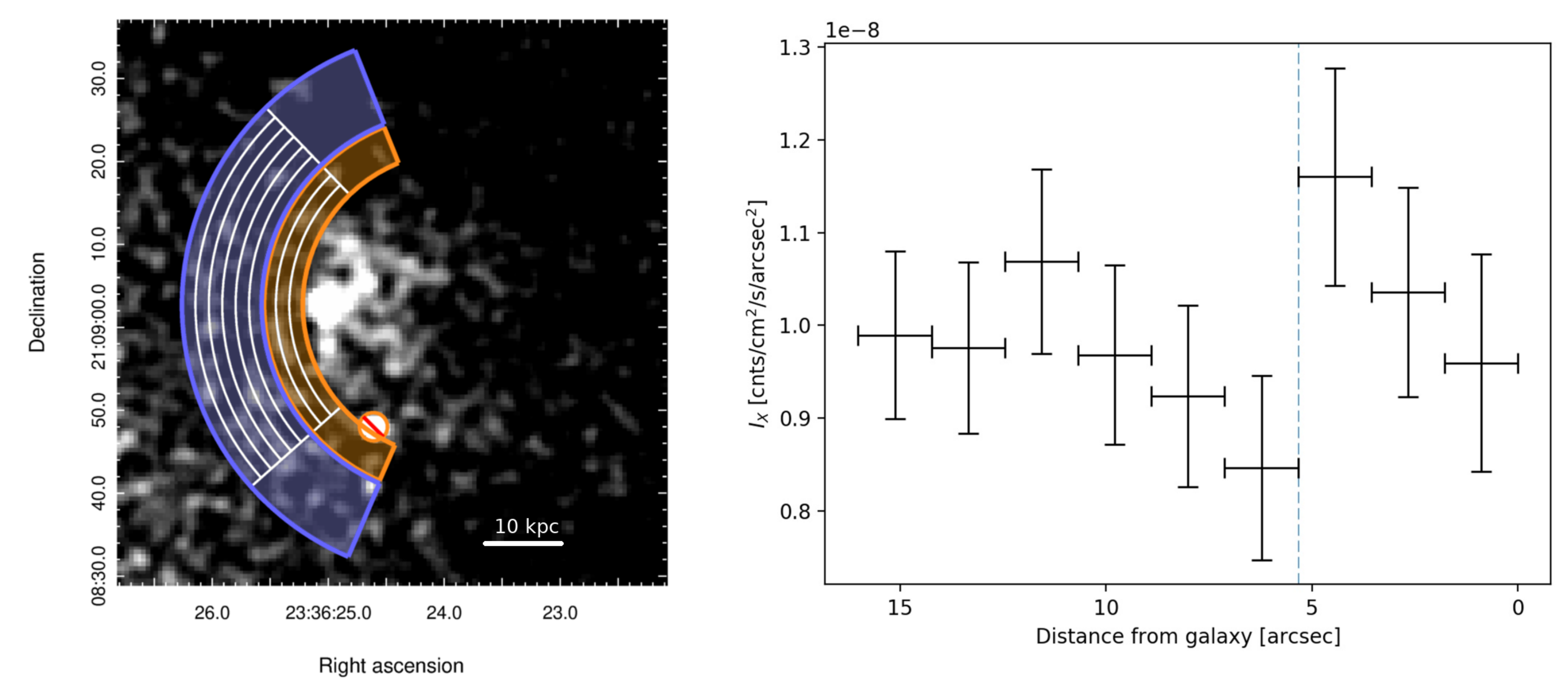}}
\caption{{\it Chandra} image of JW100 in the 0.5-2.0 keV band (left), with the sectors in which we extracted the surface brightness profile (white) and the spectra (blue and orange). The surface brightness profile (right) is taken from the galaxy to the cluster center. Note that the x-axis is inverted to match the pattern of the white sectors in the left panel. The vertical dashed line points out the location of the promising discontinuity, which is located between the orange and the blue sectors.} 
\end{figure*}

\section{Results: The spatial distribution and the physical origin of the emission at different wavelengths}


\begin{figure*}
\vspace{-2cm}
\centerline{\includegraphics[width=3.5in]{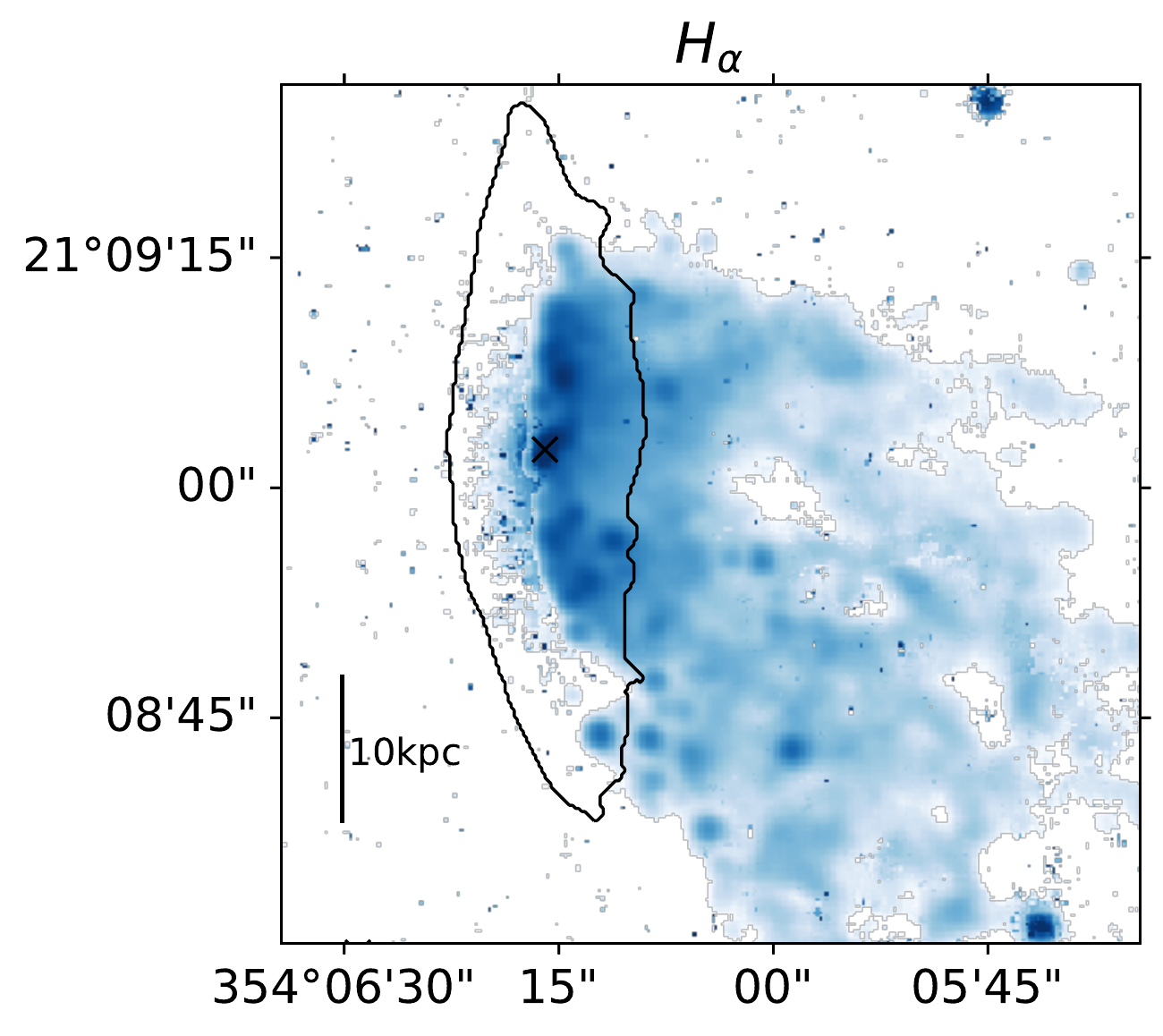}\includegraphics[width=3.5in]{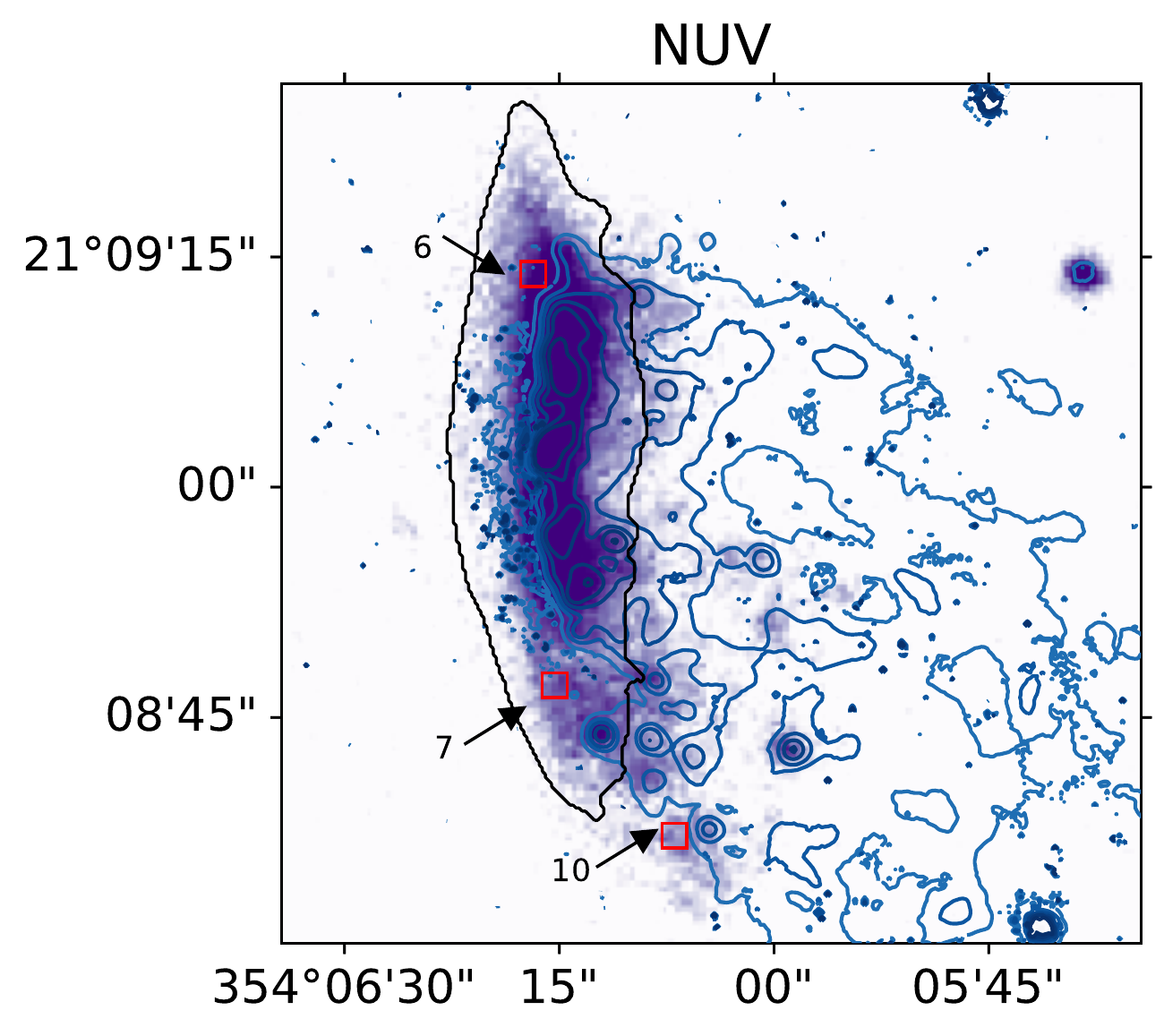}}
\centerline{\includegraphics[width=3.5in]{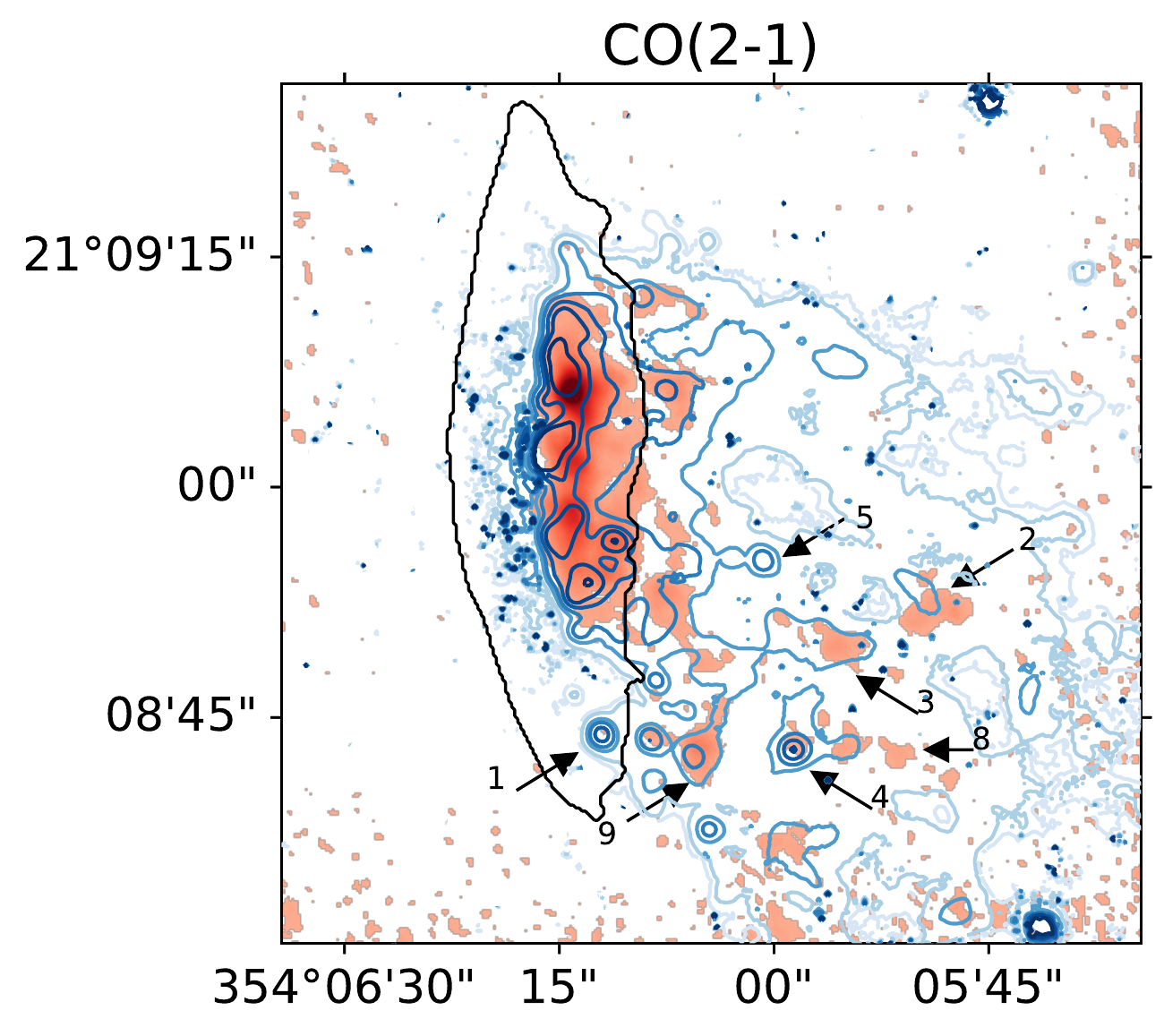}\includegraphics[width=3.5in]{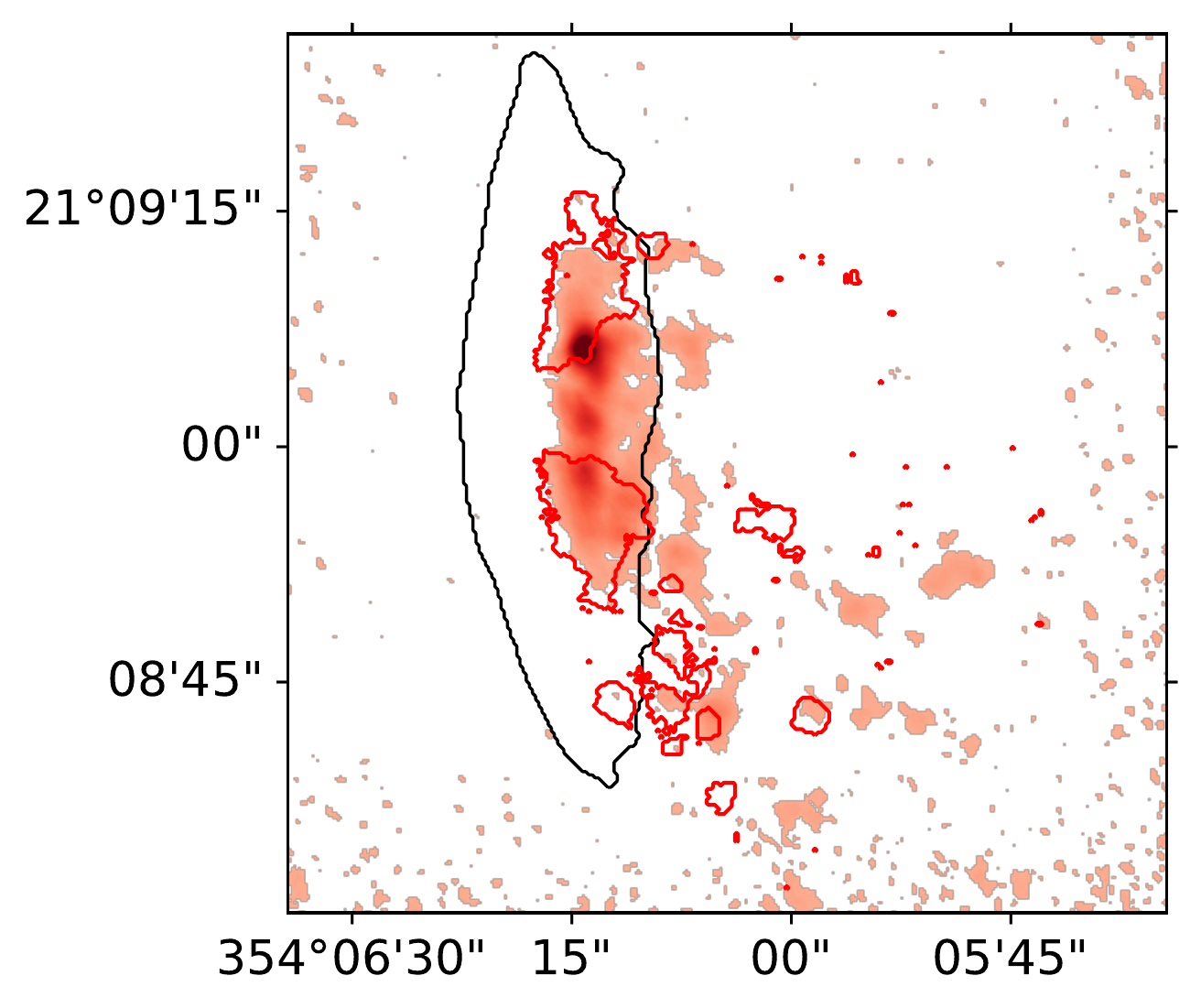}}
\centerline{\includegraphics[width=3.5in]{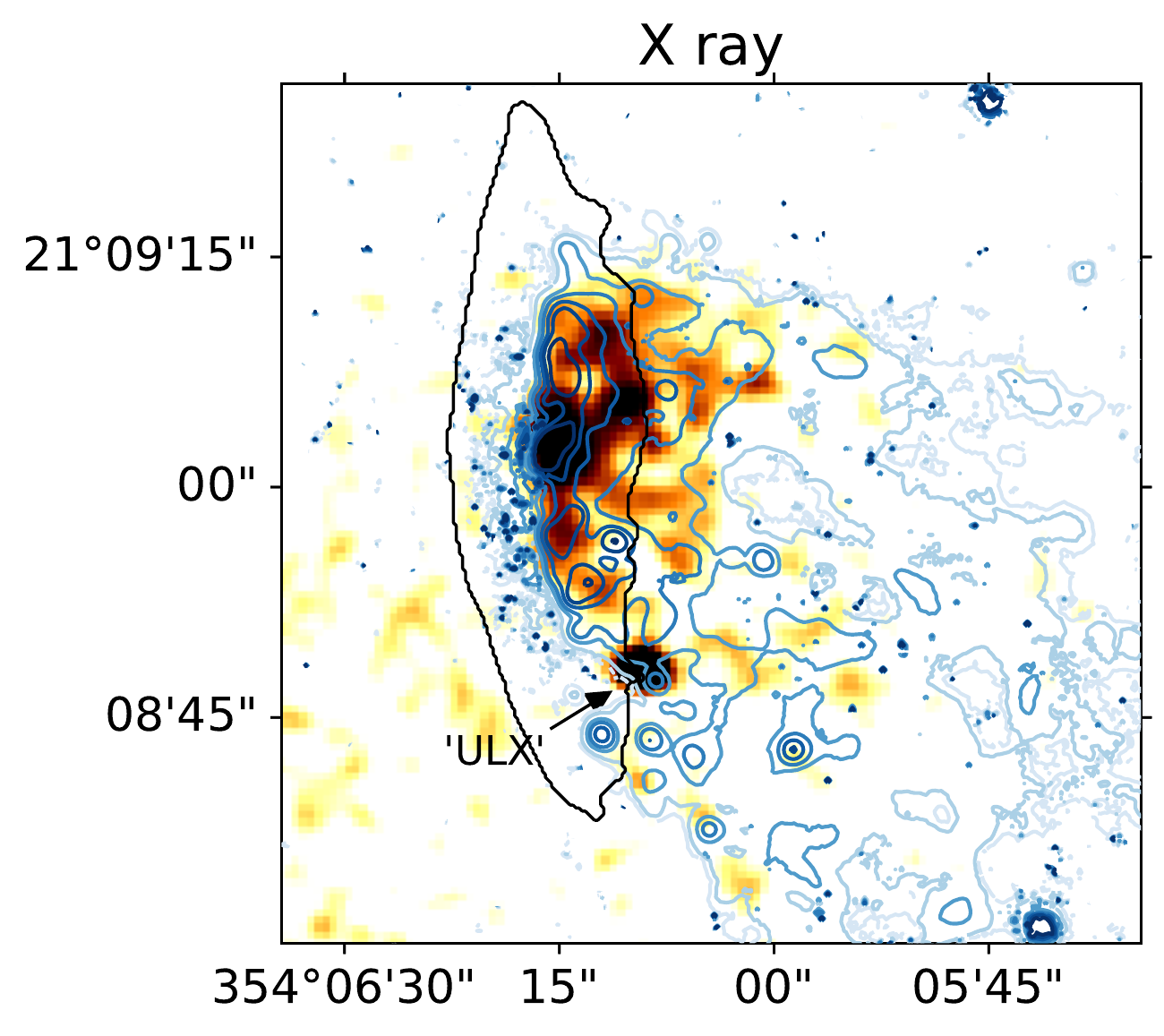}\includegraphics[width=3.5in]{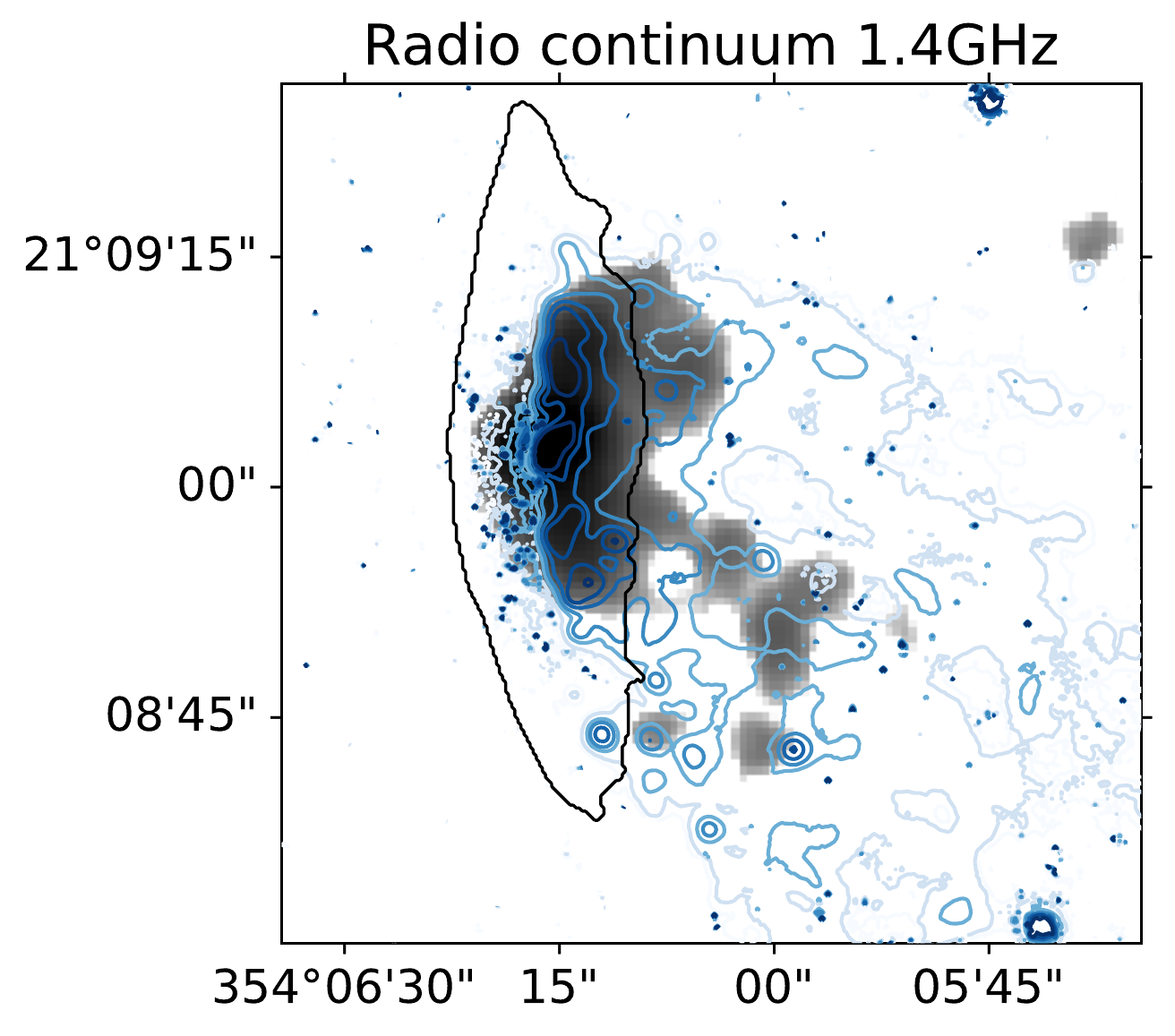}}
\caption{JW100 multi-wavelength tails. Images displayed are the MUSE
 $\rm H\alpha$ (1" resolution) and the UVIT NUV (1.2") (top panels), ALMA
CO(2-1) (1")  (middle), Chandra X-ray (0.5")  and VLA 1.4 GHz (3.8"$\times$3.4")
(bottom). Overlaid blue contours are the MUSE $\rm H\alpha$
emission. The red contours in the middle right panel outline the
star-forming regions in Fig.~2. The stellar disk region is shown by
the black contours. The positions marked with arrows and numbers are discussed in the text (\S4.1). The cross in the top left panel identifies the galaxy center, defined as the centroid of the continuum (stellar) emission underlying the $\rm H\alpha$ line from MUSE. 
} 
\end{figure*}

In this section we analyze the spatial distribution of the emission at different wavelengths.
Figure~7 presents the $\rm H\alpha$, NUV, 
CO(2-1), X-ray (0.5-2.0 keV) and radio continuum 1.4 GHz
images.
At all wavelengths, JW100 displays an extraplanar tail to the west of
the disk. Analyzing the morphology and the characteristics of the
emission at different wavelengths, in the following we investigate the physical origin of the multi-phase tail.

\subsection{$\rm H\alpha$, CO and UV}

The most extended 
tail observed is the $\rm H\alpha$ one, reaching out to at least
50kpc outside of the galaxy stellar disk, where the MUSE field-of-view ends. 
Figure~7 clearly
shows that the $\rm H\alpha$
emission is composed of bright clumps embedded in diffuse emission
\citep[see also][]{Poggianti2019}.
Within the disk, the $\rm H\alpha$-emitting gas is only present in the
western half (i.e. downstream) of the disk and out to about 14kpc from the galaxy
center along the disk major axis. At the eastern edge of the $\rm H\alpha$
emission the gas is compressed by the ram pressure (see also contours in the other panels of Fig.~7).\footnote{While shear can also remove gas from a galaxy, compression is more likely than shear to cause an enhancement in Halpha emission.  Also, as this galaxy has a significant velocity component moving towards the east, the ICM stagnation point is likely on the eastern side, reducing the strength of shear instabilities. 
}
The outer regions of the disk ($r>14 \rm kpc$ North and South of the galaxy center) and all the eastern
projected side have been already stripped of gas by ram pressure. 

The $\rm H\alpha$ velocity map (Fig.~1 in \citealt{Poggianti2017b}, not
shown here) indicates that, as is typical of jellyfish galaxies, the stripped gas
maintains the disk rotation quite coherently downstream and suggests
that in the plane of the sky the galaxy is moving with respect to the
ICM $\sim 45$ degrees North-East.

The ionization source of the bright $\rm H\alpha$
clumps is photoionization by young massive stars, thus 
star formation taking place during
the last $\leq 10^7 \rm yr$, as consistently
found by both the [SII] and the [OI] diagnostic diagrams
(Fig.~2). The star-forming clumps are mostly located in the southern part of the tail.

The origin of the ionization of the diffuse component is instead more 
uncertain, as star formation dominates according to the [SII] diagram,
while LINER-like emission dominates for the [OI] diagram (Fig.~2). These
apparently contrasting conclusions probably indicate that both star formation 
and another source of ionization contribute to
the diffuse line emission, but assessing the relative
contribution of the two processes is very hard based on diagnostic diagrams.
In the hypothesis that the stellar photons ionizing the diffuse gas
are those escaped from the HII regions within the clumps,
\citet{Poggianti2019} derived for JW100 an escape fraction of 52\%, by
far the highest in the GASP sample whose average is 18\%. Moreover, JW100 is the galaxy that most deviates
from the anticorrelation between SFR in the tail and fraction of tail
$\rm H\alpha$ emission that is diffuse (Fig.~12 in
\citealt{Poggianti2019}), demonstrating an excess of tail diffuse
emission compared to clump emission. This
is all consistent with the fact that the JW100 tail might have an
unusually high contribution from sources of ionization other than
in-situ star formation. This also agrees with the fact that, 
as visually assessed from
Fig.~7, significant UV emission is lacking in the regions of diffuse $\rm H\alpha$ emission with high [OI]/$\rm H\alpha$ ratio, supporting the notion that in situ star formation may be lacking in such areas, though we cannot exclude that fainter UV emission below our detection limit is present. A strong UV emission obscured by a large amount of dust can be excluded in the $\rm H\alpha$ diffuse emission regions, based on the moderate to low levels of dust extinction ($A_V$ values typically between 1 and 0.2 mag) derived from the Balmer decrement map observed with MUSE (not shown).

Molecular gas, as traced by CO(2-1) emission, is present only in the area
of the disk where also $\rm H\alpha$ emission is present (middle
panels in Fig.~7). Extraplanar CO complexes are also found close to the
disk (within a few kpc) in the north part of the tail, and out to $\sim$30kpc from
the disk in the southern part of the tail. A more detailed analysis of the ALMA data, both CO(2-1) and CO(1-0),
and of the spatially resolved star formation efficiency in JW100 is 
presented in a separate paper (Moretti et al. submitted) (see also \cite{Lee2017} for a comparison of CO, $\rm H\alpha$ and UV data of four ram pressure stripped galaxies in Virgo, though only within or just outside of the disks). 

For the purposes of this paper, it is interesting to compare the CO emission with both the
$\rm H\alpha$ emission (middle left panel) and with the regions that are powered by star
formation according to both the [OI] and the [SII] diagram (middle right panel).
Some of the CO complexes spatially coincide
with some of the $\rm H\alpha$ clumps, but there are also $\rm
H\alpha$ clumps with little or no CO (e.g. clump \#1 in Fig.~7) and CO complexes with only
diffuse, low surface brightness $\rm H\alpha$ emission (e.g. clump \#2). Moreover, not all the
$\rm H\alpha$ clumps with corresponding CO emission are classified as
star-forming according to the [OI] diagram (e.g. clump \#3). 

These findings can be reasonably explained by a combination of two factors.
The fact that not all $\rm H\alpha$ clumps with 
CO emission are classified as star-forming according to the [OI]
diagram, might be due to the coexistence of different ionization
mechanisms contributing at the same location, or at least powering
ionized regions superimposed along the same line of sight.

The fact that CO and $\rm H\alpha$ clumps do not always spatially coincide
might be due instead to
the evolutionary stage of the star-forming
regions.
Each star-forming 
region will go through four phases: a) a molecular-gas-only
phase (no massive stars formed yet); b) a molecular-gas+ionized gas+UV
light phase (massive stars have had the time to ionize the surrounding
gas and they shine in the UV, but they have not dispersed the
remaining molecular cloud yet); c)  a phase with ionized gas+UV light (the
molecular clouds have been destroyed, there are still massive stars that
ionize the gas and shine), and d) a UV-light-only phase. In the latter phase 
the stars more massive
than about 20 $M_{\odot}$ have died and the ionizing radiation of the
stars left is not capable to ionize a significant amount of gas, but
the UV radiation from the most massive stars on the main sequence is
still sufficient to let the region shine in the UV. 
This is the case if the SF occurred more than $10^7$ yr, and less than a few times $10^8$ yr ago.

Comparing the first four panels of
Fig.~7, we see this evolutionary sequence of star-forming regions,
going from regions like \#2, \#3 and \#8  (CO clouds but no
bright $\rm H\alpha$ clump nor UV yet), to regions like \#4 and \#9 (CO+$\rm
H\alpha$+UV), to regions like \#1 and \#5 ($\rm H\alpha$+UV, very little or
no CO left) and, finally, to UV-only regions 
such as region \#10 in the tail and regions \#6 and \#7 in the outer parts of the disk, where only the UV light has
remained to testify the recently quenched star formation where all gas
has been stripped. This decoupling of the various phases of the  star formation process is similar to the decoupling observed by e.g. \citet{Kruijssen2019}, who have found a decoupling of CO-dominated clumps and $\rm H\alpha$-dominated clumps on $\sim 100$pc scales in NGC300, a close face-on star-forming disk galaxy.

Interestingly, in the tail of JW100, we see that the different phases are located according to a spatial progression,
going from left
to right (oldest to youngest) in the southern part of the tail.
This spatial progression is summarized in Fig.~8, where the different stages are shown in different colors: red where only CO is observed, orange for CO+UV+$\rm H\alpha$, green for UV+$\rm H\alpha$, and cyan for UV-only. Here only the $\rm H\alpha$ emission due to SF according to the [OI] diagram has been considered. 
The first stage of the star formation sequence (only CO) is preferentially located to the right of the other colors, and the following stages are found progressively to the left. In some cases, this sequence is at least partially observed even {\it within} an individual star-forming clump (e.g. \#4, from right to left: orange, green and cyan). This latter effect is similar to the "fireballs" observed in the tail of ESO137-001 by \citet{Jachym2019}.

The progression observed in Fig.~8 thus traces the timing of star formation in the tail, and
strongly suggests that the molecular clouds further away from the disk
have not formed stars (yet). This agrees with the (counter-intuitive) stellar age gradient in the tail found in some jellyfish galaxies, by which younger stellar clumps are found further away from the disk (e.g. IC3418 \citet{Fumagalli2011,Kenney2004}, RB199 \citet{Yoshida2008}, JO201 \citet{Bellhouse2019}). Hydrodynamical simulations do predict very recent star formation preferentially further out in the tail than closer in the disk for some infalling angles (e.g. Fig.~1 in \citealt{Roediger2014} for a face-on infall). A detailed comparison with simulations is beyond the scope of this paper, and will be the subject of a future work.

The recently quenched star formation occurring where the gas has been
totally stripped 
is testified not only by the UV emission, but also from the strong 
Balmer lines in absorption (accompanied by the lack of emission lines) in the MUSE 
spectra. As an example, the spectra of regions \#6 and \#7 shown in
Fig.~9 display the strong Balmer lines typical of post-starforming and 
post-starburst regions \citep{Poggianti1999}. The $\rm H\beta$ rest
frame equivalent width of these spectra is 6.7 and 6.6 \AA,
respectively, which can only arise from stellar populations less than
1 Gyr old whose spectrum is dominated by A-type stars.

\begin{figure}
\centerline{\includegraphics[width=3.5in]{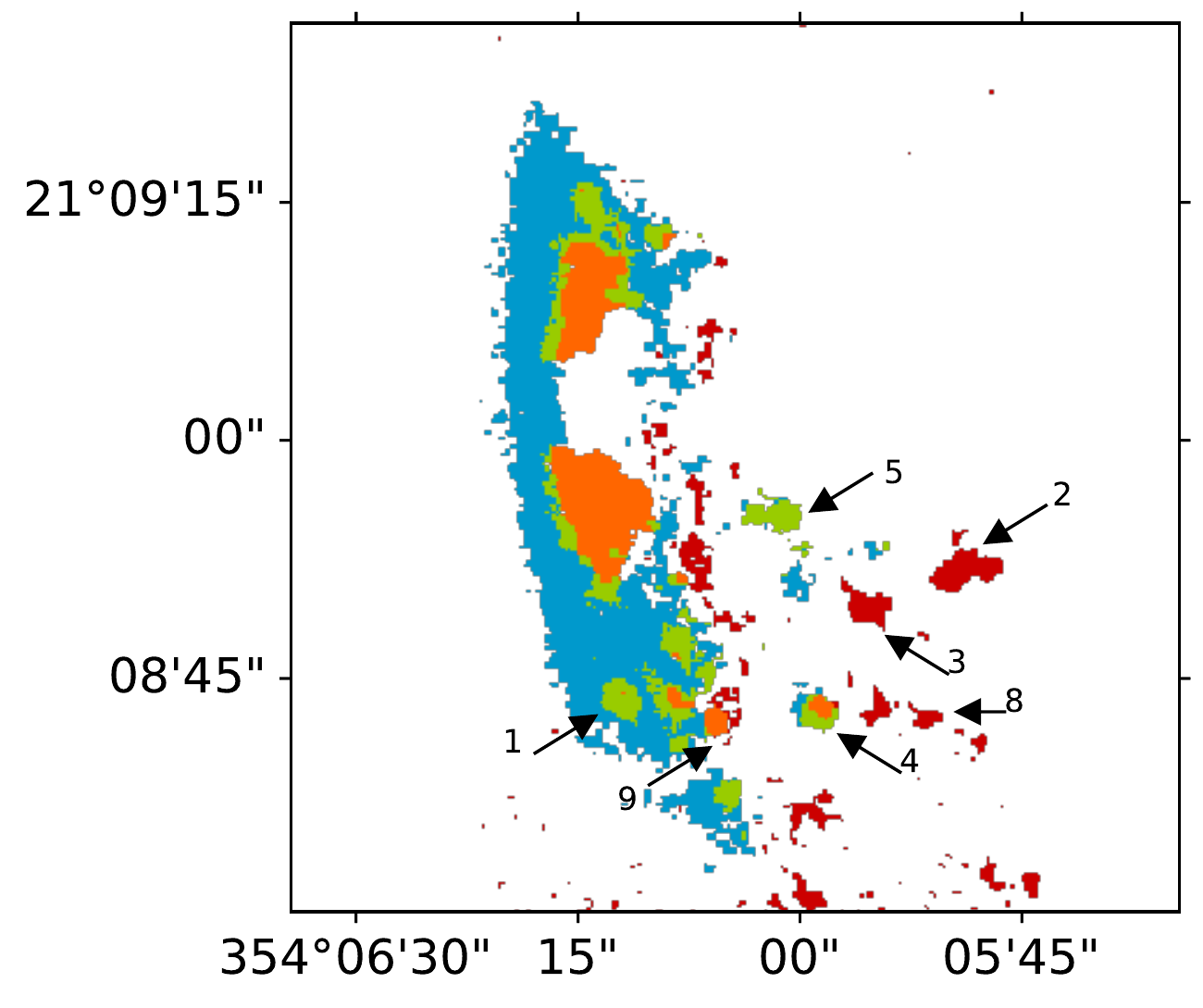}}
\caption{The star formation sequence created by ram pressure: only CO
  (red), CO+SF-powered $\rm H\alpha$+UV (orange), SF-powered $\rm
  H\alpha$ + UV (green), and only UV (cyan). The regions with
  SF-powered $\rm H\alpha$ are star-forming also according to the [OI] line (right panel in Fig.~2). Numbers as in the middle left panel of Fig.~6.
}
\end{figure}

To summarize, the observables that are more closely linked with ongoing/recent star
formation (the gas ionized by star formation, the molecular gas and
the UV emission) all point to the southern half of the tail as the
location of in-situ star formation extending much further away from the disk than in the northern part. Although overall these three observables depict a similar picture,
the exact location of $\rm H\alpha$, CO and UV emission does not
always coincide on small scales. This mismatch can be ascribed to the
different star formation stages and timescales traced by $\rm
H\alpha$, CO and UV (thus an evolutionary sequence within star forming
regions), in some cases combined with the contribution of an
additional source of gas ionization that gives rise to peculiar
optical line ratios observed with MUSE, such as shock heating, thermal conduction from the ICM or mixing of the ISM and the ICM (see \S5.3). 

\begin{figure*}
\centerline{\includegraphics[width=3.5in]{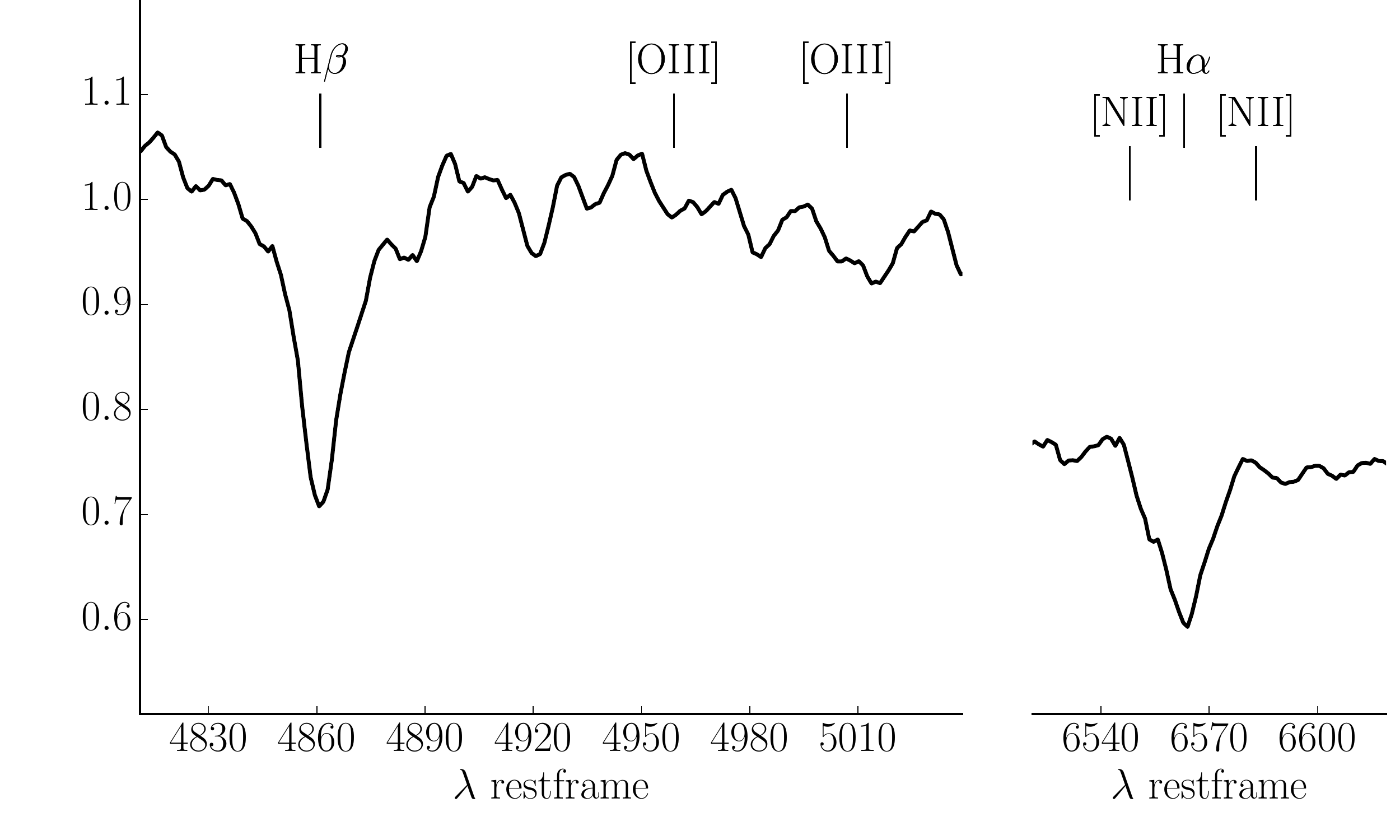}\includegraphics[width=3.5in]{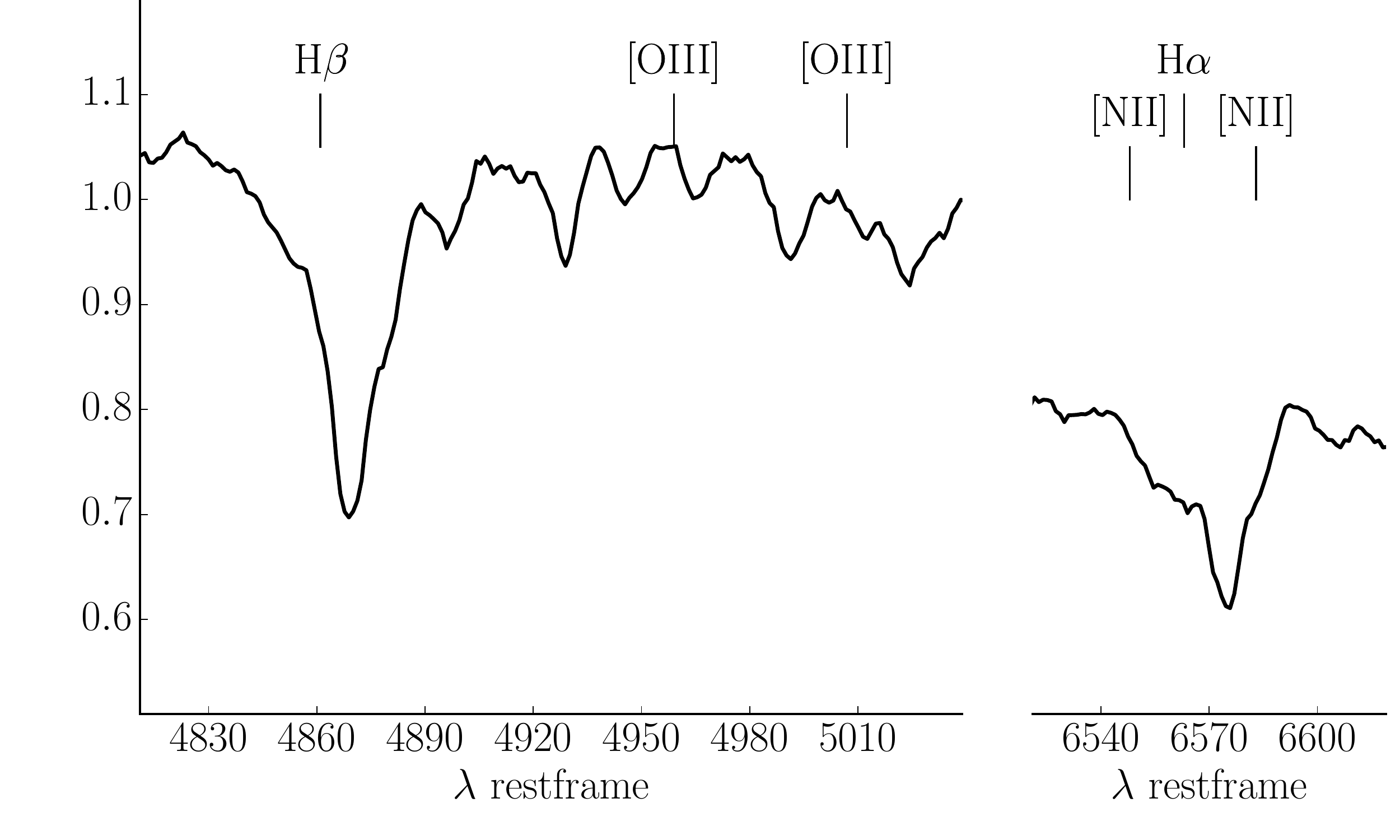}}
\caption{Spectra of regions (6) (left) and (7) (right), see red
 squares in Fig.~7. 
}
\end{figure*}

\subsection{Radio-continuum}


Although care should be taken when comparing images at different resolutions (given the amount of tapering required to better map the extended/diffuse radio structure, the resolution of the 1.4 GHz map is a factor $\sim 3$ worse than that of the other images, which is about 1 arcsec), we note that 
the 1.4 GHz emission appears to have a different spatial distribution from 
all other wavelengths (bottom right panel in Fig.~7).

In the disk, it globally
coincides with the location of the $\rm H\alpha$ and CO (and X-ray)
emission, 
though with a slight extension to the east of the ram
pressure edge, 
in correspondance to the galaxy center, most likely due
to the radio continuum emission from the AGN, whose point-like
morphology is clearly detected in the high-resolution images at 5 GHz presented in \citet[][Fig.~4]{Gitti2013} and \citet[][Fig.~1]{Ignesti2017}.

Outside of the disk, 
extended 1.4 GHz
emission is detected. We note that such emission is indeed diffuse as it is not evident in the 1.4 GHz maps which have a
synthesized beam about or slightly higher than 1 arcsec \citep[the untapered reconstructed images, see Figs. 1 and 2 of][]{Gitti2013}. In particular,
a northern radio continuum tail extends
approximately as much as the NUV and X-ray emission. There is also
a radio south tail which covers patchy areas with no CO, no bright $\rm
H\alpha$ knots (except for one knot in the south-east), and
almost no X-ray detected.
The radio continuum emission therefore is 
generally present in regions that are
lacking 
ongoing clumpy star formation/high molecular gas content. There is instead at least some UV emission in the majority of radio-emitting tail regions, except for the south-west
radio clump.


In order to probe the thermal or non-thermal origin of the radio 
emission, we estimated the spectral index of the tail between 1.4 and 
5.5 GHz using the VLA maps presented in Ignesti et al. (2017, Fig. 1). 
At 5.5 GHz we do not observe diffuse radio emission in the tail, so 
by considering the 3$\sigma$ level, we could estimate only a lower limit 
for the spectral index $\alpha>0.5$ (in this work the spectral index 
$\alpha$ is defined such as $S\propto \nu^{-\alpha}$, where $S$ is the 
radio flux and $\nu$ is the frequency).
\citet{Tabatabaei2016}
fitted the 1-10 GHz spectral energy distribution of nearby galaxies using 
a Bayesian Markov Chain Monte Carlo technique in order to disentangle 
the thermal and non-thermal contributions to the radio emission. These 
authors measured a total spectral index (combination of thermal and 
non-thermal components) ranging from 0.5 to 1.0. Then, by assuming 
$\alpha=0.1$ for the thermal radio emission, they estimated a mean 
thermal fraction at 1.4 GHz of 10-13$\%$. Therefore, from our estimated 
lower limit  we may conclude that the fraction of non-thermal 
contribution to the total radio emission from the tail of JW100 is more
than 90$\%$, thus indicating that we are mainly observing synchrotron 
emission of relativistic electrons diffused in the tail of JW100. 
The spectral index of the disk is $\sim$0.6-0.7, and it 
may be contaminated by the AGN.

The origin of relativistic electrons and magnetic field in the stripped tail is uncertain. The expected $L_{1.4GHz}$ associated with the SFR derived from MUSE and converted from a Salpeter to a Chabrier IMF using the \cite{Bell2003} calibration SFR=$3.25 \times 10^{-22} L_{1.4GHz} 
 (\rm W \, Hz^{-1})$  is 2.5$\cdot10^{21}$ W Hz$^{-1}$ and 9.8$\cdot10^{21}$ W Hz$^{-1}$ for the tail and the disk, respectively. These values are consistent with the observed $L_{1.4GHz}$ (Tab. 5), thus suggesting that core-collapse supernovae may be the dominant source of relativistic electrons. However, we can not exclude a contribution of other sources, such as stripping of relativistic electrons from the galaxy due to ICM winds \citep{Murphy2009}. Regarding the magnetic field in the tail, it may be the combination of the ISM magnetic field following the stripped plasma and the ICM magnetic field, but with the present data we could not perform a study of polarized emission to disentangle them. The magnetic field measurement in a GASP jellyfish tail will be presented in Mueller et al. (2019 in prep.). Interestingly, the radio emission in the tail of JW100 on small scales
often does not {\sl spatially} coincide with the regions of ongoing
star formation (the brightest $\rm H\alpha$ clumps and the CO
clumps). Assuming that most of relativistic electrons derive from SFR, this might be due to timescale issues. During their lifetime
the accelerated electrons can travel large distances, and/or 
at the location of the youngest star-forming regions possibly there aren't yet
powerful accelerating sources of electrons, i.e. there have not been
supernova explosions.

\subsection{X-ray}


In contrast with the star formation tracers described in \S5.1, the X-ray extraplanar emission is much more conspicuous in the northern half of the tail than in the southern half, with a ratio in the 0.5-2.0 keV band between north and south of $\sim4$ in luminosity and $\sim6.5$ in surface brightness  (see  regions \# 1-2 vs. \# 3-4 in Fig. 10).
The southern tail contains a bright point source, a candidate Ultra Luminous X-ray source (ULX), that is located just
outside of the stellar disk, which will be discussed in detail in
\S6. 

Interestingly, the sharp eastern edges of X-ray, $\rm H\alpha$ and CO emission
in the disk coincide, along a vertical line where all three gas phases
are compressed by the ram pressure. The extension and location of 
all three gas phases in the disk also coincide, as they are only found
in the western side of the central 14 kpc, while all the outer disk regions and all the eastern side of the disk are devoid of all gas phases.

JW100 shows its own extended X-ray emission, on a scale $\geq$ a dozen kpc, standing above the emission of the cluster (see spectral analysis in \S4.2) . This emission might have different origins: a) SF within the tail; b) stripped hot galaxy halo; c) heating of the cold ISM, either by shocks, thermal conduction or mixing with the ICM. We do not see instead any anisotropic features in the X--ray emission that may suggest the presence of jets powered by AGN; also unlikely is a contribution to the extended X-ray emission from the nuclear outflow revealed by optical emission lines, considering the low mass rate (see \S2 and \S6).

a) Let's start considering the first hypothesis: 
in the presence of SF, the dominant contribution to the X-ray emission is expected from high-mass X-ray binaries, that have a lifetime of a few $10^7$ yr and dominate over the emission of low-mass X-ray binaries when there is vigorous ongoing SF. A smaller contribution arises from the hot ISM ionized by supernovae and massive stars. Each of these contributions, and the sum of the two, correlate well with other SFR indicators \citep{Ranalli2003, Mineo2012a,Mineo2012b,Mineo2014}.

To test whether the observed X-ray luminosity of JW100 is compatible with the SFR measured from the optical lines, we use the $L_X-SFR$
calibration from \citet{Mineo2014} converted from a Salpeter to a 
Chabrier IMF and from 
0.5-8keV to 0.5-10keV assuming a factor 1.11:
$SFR=1.32 \times 10^{-40} L_{X(0.5-10)} \rm \, erg \, s^{-1}$.
With this calibration, the 
total X-ray luminosity of JW100 listed in Table~4 corresponds to 
SFR=28$\pm 5$ $M_{\odot} \, yr^{-1}$ (model apec+apec), 
SFR=60$\pm 11$ (model apec+cemekl with $T_{max}$ fixed), or 
SFR=33$\pm 6$ $M_{\odot} \, yr^{-1}$ (model apec+cemekl without
$T_{max}$ fixed). We remind the reader that X-ray point sources (AGN
and candidate ULX) have already been excluded from the calculation of
the X-ray luminosities. Even excluding the region of the disk where
the X-ray and $\rm H\alpha$ contours are compressed by ram pressure
(identified from Fig.~7), where the X-ray luminosity could be boosted,
the derived SFR would still be very high, ranging between 21$\pm 4$ and 46$\pm 9$ $M_{\odot} \, yr^{-1}$.
In the range 0.5-10keV, $\sim 50$\% of the counts come from the tail and assuming the shape of the spectrum in the tail is similar to the total one this 
should correspond to a tail SFR between $\sim$14$\pm 3$ and $\sim$30$\pm 7$ $M_{\odot} \, yr^{-1}$
depending on the X-ray model adopted. 

The X-ray-based SFR values are much higher than 
those measured from the 
dust-corrected, absorption corrected $\rm H\alpha$ luminosity ($4
M_{\odot} \, yr^{-1}$ total,$\sim 1 M_{\odot} \, yr^{-1}$ in the tail, see \S2) even under the most generous assumptions (using all the regions powered by star formation according to the [SII] diagram). The scatter in the $L_X - SFR$ relation in \citet{Mineo2014} is not able to account for the low $\rm H\alpha$-based SFRs, which  a factor between 6 and 30 lower than the X-ray-based values. Using the 0.5-2 keV $L_X-SFR$ relation from \citet{Ranalli2003} transformed from a Salpeter to a Chabrier IMF yields slightly lower X-ray-based SFR (between 14 and 24 $\sim 1 M_{\odot} \, yr^{-1}$) than with the Mineo calibration, but these values are still 
higher than the $\rm H\alpha$-based SFR by a factor between 4 and 6.

It is interesting to note that comparing the total X-ray emission (having excluded the AGN, Table~4) and the 1.4GHz emission (Table~5), the X-ray emission of JW100 is more than an order of magnitude too high for its radio continuum emission, according to the relation shown in Fig.~1 of \citet{Mineo2014}. Thus, while the radio continuum and the $\rm H\alpha$ emission are consistent and can be explained with star formation (see \S5.2), the X-ray emission has a significant excess both with respect to $\rm H\alpha$ and the radio continuum.

We conclude that 
it is necessary to invoke an additional source of X-ray emission other than 
the sources linked with ongoing star formation.



b) The second hypothesis concerns the galaxy hot, X-ray emitting halo (see \cite{Bregman2018} for a review of hot circumgalactic gas). Before being stripped, we may expect JW100 to have possessed a hot, X-ray-emitting halo,
that could have been stretched during the
stripping, producing the observed extra-planar diffuse emission 
\citep{Bekki2009}. 
The galaxy NGC1961, whose stellar mass is identical to that of
JW100, has a hot halo with an X-ray luminosity within 50 projected kpc
$L_X(0.5-2) = 8.9 \pm 1.2 \times 10^{40} \rm \, erg \, s^{-1}$ \citep{Anderson2016}, which is 
not too dissimilar from the $L_X(0.5-2)$ of JW100, that is between 2 and 3.5
$\times 10^{41} \rm \, erg \, s^{-1}$ (Table~4). Similar X-ray
luminosites ($L_X(0.5-2) = 5-10 \times 10^{40} \rm \, erg \, s^{-1}$) have been measured for the inner $\sim 40$kpc hot halo of
other few galaxies as massive as JW100 (Fig.~4 in \citealt{Li2016}).
Given the expected enhancement of X-ray luminosity in the region compressed by ram pressure, in principle the JW100 X-ray luminosity is compatible with being the stripped hot galaxy halo.

An easy way to evaluate the importance of the hot halo stripping is to compare the ram pressure, $p_{\rm ram} \sim 3\times 10^{-10}$ dyne cm$^{-2}$ (see Appendix) to the halo thermal pressure, $p_{\rm halo} = n_{\rm halo} k T_{\rm halo}$, where the halo temperature is assumed to be $T_{\rm halo}\sim 6\times 10^6$ K (Appendix). Stripping is efficient where $p_{\rm ram}\geq p_{\rm halo}$, or where $n_{\rm halo} \lesssim 0.3$ cm$^{-3}$. The latter condition is likely to be verified everywhere but, perhaps, the central kpc or so (e.g. \cite{Bregman2018}).
We also tested the hot halo stripping hypothesis with a simple model adopting a mass model for the galaxy including a stellar bulge, a stellar disk and a dark matter halo (see Appendix for details). 
This model finds that the hot halo will be stripped down to the galactic central kpc or less on very short timescales ($\sim 15$ Myr), which makes the stripped halo hypothesis for the observed X-ray halo highly unlikely.\footnote{An example of hot halo/intragroup medium stripping from an early-type galaxy could be CGCG254-021
which is the very massive, brightest galaxy in the Z8338 cluster, in which the X-ray gas is completely detached from the galaxy \citep{Schellenberger2015}.}




c) Finally, the X-ray emission of JW100 can arise from heating of the ISM either due to
shock heating, thermal conduction from the ICM, mixing of the ISM and the ICM, or cooling of the ICM onto the colder stripped ISM. The latter scenario can be explored by considering the Field length for a static cold cloud immersed in a hot medium, $\lambda_F = [\kappa (T) T/n^2\Lambda(T)]^{1/2}$, where $\kappa$ is the thermal conductivity, $T$ and $n$ are the temperature and number density of the hot gas and $\Lambda$ is the cooling function. 
This is a measure of the balance between the cloud energy gain by conduction and the energy loss by radiation (see \citet{Mckee1990}). For a cloud size larger that $\lambda_F$, radiative cooling dominates and the hot ICM condenses on the cloud. Numerically,  $\lambda_F \sim 136 \phi_c^{1/2} T_{e,7}^{7/4} n_e^{-1} \Lambda_{-23}^{-1/2} \; {\rm pc} \approx 450$ kpc for the ICM surrounding JW100 \citep{Mckee1990}. For this estimate we have adopted $n\sim 3\times 10^{-3}$ cm$^{-3}$ and $T\sim 3.5$ keV from the X-ray analysis above. The factor $\phi_c\le 1$, which describes the suppression of the conductivity in a magnetized plasma, has been set $\sim 1$. The value for $\lambda_F$, much larger than the size of the JW100 cold ISM, makes the cooling scenario unpalatable, unless $\phi_c\ll1$.


Mixing (and heating) of the stripped cold ISM with the hot ICM was invoked to explain the X-ray tail of the jellyfish ESO-137-001 in the Abell 3627 cluster: the X-ray would arise from
a warm contact surface 
between the ISM and the ICM which is emitting in X-rays \citep{Sun2010}.



In an oversimplified picture, the two types of X-ray spectral fitting models 
discussed in {\bf \S4.2} might represent the scenarios b) and c): the {\tt apec+apec} model would correspond to the stripped corona hypothesis, while the {\tt apec+cemekl} model to the ISM heating hypothesis.
Unfortunately, these two models are statistically
indistinguishable (see ${\chi}^2$ and DOF values in Table~4).
Therefore, what we can conclude from our X-ray spectral modeling of JW100 (\S4.2 and Table~4) is that the data are consistent with a multi-temperature gas, thus with the ISM heating scenario, but can neither rule out the hot halo hypothesis nor discriminate among the possible heating mechanisms. 

The X-ray plasma metallicity would be a key probe to
discern the models, because in
the stripped hot corona scenario we would expect the metallicity to be lower ($Z\simeq0.1-0.3$ $Z_{\odot}$, Werner et al. 2019) than that of the heated stripped gas, whose MUSE metallicity is solar and supersolar (Franchetto et al. in prep.). However, the existing X-ray data do not allow us to estimate the metallicity of the X-ray emitting gas, and longer {\it Chandra} exposures are needed.

\subsubsection{The heating ISM scenario and the $\rm H\alpha$-X-ray correlation}

The fact that the morphology of the X-ray gas in the disk follows
exactly the morphology of the $\rm H \alpha$ gas is consistent with the ISM
heating scenario. We also note that in simulations with an angled wind
we find that once initial stripping has occurred, dense clouds tend to
be found on the side of the tail that is more downwind (Tonnesen et
al. in prep.).  While much of the X-ray emission is near stripped
dense clouds in these simulations, there are times when the brightest
X-ray emission 
is found near the disk and closer to the upwind side of the disk than the dense clouds, as in our JW100 observations.    
Moreover, in the simulations by \citet{Tonnesen2011} the authors can reproduce the bright $\rm H \alpha$ and X-ray emission of the galaxy ESO 137-001, and argue that bright X-ray emission occurs when the stripped ISM is heated and mixed into a high-pressure ICM (greater than $\sim 9 \times 10^{-12} \rm \, erg \, cm^{-3}$). The JW100 ICM pressure of $\sim 3 \times 10^{-11} \rm \, erg \, cm^{-3}$ is above this threshold, and about twice the pressure around ESO137-001 (see \S5.3.2 for a comparison of the two galaxies). 


To further investigate the heated ISM hypothesis, we plot in
Fig.~10 the $\rm H\alpha$ surface brightness (corrected both for stellar absorption and dust extinction computed from the Balmer decrement) versus the X-ray surface brightness for the regions identified in Fig.~2. 
Some of these regions coincide with star-forming regions according to both the [OI] and the [SII] MUSE diagnostic diagrams (red points), while others correspond to regions of [OI]-LINER-like emission (blue points), which are mostly classified as star-forming by [SII].\footnote{We have tried sampling smaller 
 regions within the largest regions, results do not change and the 
 relation for the diffuse component persists.} The former  generally have a higher average $\rm H\alpha$ surface brightness than the latter, due to the star-forming clumps.

To estimate the X-ray surface brightness of each region, we extracted the corresponding spectra by excluding the cluster contribution and, then, we fitted the spectra with an absorbed {\tt apec} model. Due to the low statistics, we could not estimate the local properties of each region, so, under the assumption that the properties of the X-ray emitting plasma are the same all over the galaxy, we fixed the temperature and the metallicity to the values that we estimated for the whole galaxy ($kT$=0.82 keV, $Z$=1 $Z_\odot$) and we derived the luminosity in the 0.5-8.0 keV band from the fit normalizations, that we ultimately converted in surface brightness.

Figure~10 presents some striking results.
First of all, the $\rm H\alpha$ and X-ray surface brightnesses correlate as $I_\text{$H_{\alpha}$}\propto I_\text{X}^s$, where $I_\text{$H_{\alpha}$}$ and $I_\text{X}$ are the surface brightnesses of the two bands. This result suggests a physical relation between the two emission processes. Interestingly, the correlation is different for star-forming regions and for regions with an [OI] excess, with $s=0.87\pm0.17$ and $s=0.44\pm0.17$ respectively. In both cases the Spearman correlation coefficients are $>0.9$.
At similar $\rm H\alpha$ surface brightness, [OI]-excess regions have a higher X-ray brightness. This could be consistent with ISM heating in the [OI]-excess regions, which could
account for the main observational results, explaining:
a) the existence of the correlation $\rm H\alpha$-X and of the X-ray excess compared to star-forming regions; b) the additional source of ionization/excitation of the stripped gas, causing the different optical line ratios, in particular the [OI], and c) the lack of significant molecular gas and ongoing star formation in these regions.

On the other hand, the $\rm H\alpha$-X surface brightness relation of even strongly star forming regions cannot be fully explained by the expected emission at these wavelengths from a given SFR under standard conditions. In \S5.3.0 we have discussed how on a global scale there is a significant X-ray excess for the $\rm H\alpha$ emission (SFR) observed.  
In Fig.~10 we see the same effect but on the scale of individual regions:
all the starforming regions present an excess of X-ray emission compared to the SFR calibration commonly used in the literature, represented by the solid black line. The global values (green points in Fig.~10) are dominated by the star-forming regions, especially in the disk (green square), while the overall value (green diamond) lies in between the star-forming and the diffuse emission regions.

At this point we can only speculate on the possible origin for the discrepancy between the star-forming points and the standard SFR relation line in Fig.~10:

a) The standard SFR calibrations may not apply under the extraordinary physical conditions in which stars form in the tail (IMF, different timescales probed by the two indicators etc). To our knowledge, the $\rm H\alpha$ vs X-ray-based SFR estimates have not been tested in the literature even for normal spirals, 
although given the good correlations between X-ray and radio/UV+FIR and between $\rm H\alpha$ and UV+FIR a discrepancy similar to the one we observe here would be surprising in normal spirals.

b) There could be an additional source of X-ray emission due to ISM heating even along the line of sight of star-forming regions (though its relative importance should be lower than in the [OI]-excess regions). This is clearly possible, but the effect should be conspicuous, because the observed X-ray surface brightness of star-forming regions is a factor 5-8 higher than expected from the $\rm H\alpha$ and  in order to reconcile the two SFR estimates the majority of the X-ray flux should arise from heating.
This effect might be seen from another point of view. The main, underlying relation in Figure~10 might be the one traced by the blue points, in which the emission in $\rm H\alpha$ and X arise from exactly the same process, e.g. ISM heating.
The star-forming points would lie {\it above} this relation due to an excess of $\rm H\alpha$ flux due to star formation.

To conclude, our data are consistent with
the X-ray emission of JW100 coming from warm regions that envelope the cold ISM observed in $\rm H\alpha$, due to ISM-ICM mixing or thermal heating of the ISM due to the ICM, or shock heating. It is also consistent with the [OI]-excess observed in the JW100 tail being a consequence of efficient mixing/thermal heating/shock heating.

However, this tentative hypothesis is far from being demonstrated. Confirming it or disproving it require deeper X-ray observations, and will benefit from a comparison with the $\rm H\alpha$-X-ray surface brightness relation in other environmental conditions. For example, the diffuse LINER tail of JW100 shows interesting similarities, in term of both optical and X-ray properties, with the multi-phase filaments observed at the center of clusters (e.g., \citet{Werner2013}). A fascinating explanation for that similarity might be that these filamentary structures, although on different scales, are generated by the motion of a substructure in the ICM (the galaxy, in our case, or an uplifting AGN cavity)  that triggers the phase mixing and the ICM cooling.
In addition, in principle, a powerful tracer of mixing or conductive layers are UV lines, especially [OVI], which probe gas at intermediate temperatures few$\times 10^5$ K. This gas phase is expected to be more abundant in the heating scenario, therefore [OVI] observations might help discriminating between the hypotheses.
Moreover, cloud-scale high resolution simulations of cold clumps embedded in a hot medium with realistic conditions for the stripped tails will be fundamental to assess the hypothesis proposed in this paper and analyze the various possibilities for ISM heating (e.g. \citet{Bruggen2016,Armillotta2016,Gronke2018}).

\subsubsection{Comparison with ESO137-001}


We compare the X-ray properties derived in this paper with those obtained for ESO137-001 by Sun et al. (2010), 
A comparison of the star formation and the CO properties
for JW100 and ESO137-001 are given in Poggianti et al. (2019) and Moretti et al. (2019 submitted), respectively.

We note  that the stellar mass of JW100 is almost two order of magnitudes higher than that of ESO137-001. Given the low mass of ESO137-001, this galaxy is not expected to have had a hot corona, thus its X-ray emission must come from the interaction between cold ISM and ICM. This would suggest that the same mechanism generate the X-ray emission for JW100, giving strength to our discussion above. 




The X-ray luminosity $L_{0.5-2keV}$ is 1.1$\times 10^{41} \rm \, erg \, s^{-1}$ for ESO137-001 and
about twice this value for JW100 (2-3.4$\times 10^{41}\rm \, erg \, s^{-1}$, Table~4).
The morphologies of the X-ray tails of the two galaxies are very different:
that of JW100, at the current depth of the {\it Chandra} data, is within the galaxy tidal truncation radius (truncated by the cluster potential, about 38 kpc), while the observed 80kpc X-ray tail of ESO137-001 is well outside of its truncation radius. 
However, we calculate that with the exposure time and the {\it Chandra} effective area of the Sun et al. observations (given the degrading of $\it Chandra$ performance with time)
we would have had 2.5 times the counts we observed. Given the $\rm H\alpha$-X ray correlation we observe, it is reasonable to expect the X-ray tail to be as extended as the $\rm H\alpha$ tail. Deeper X-ray observations would be needed to observe the full extent of the X-ray tail in JW100.


Interestingly, the X-ray temperature derived for the two galaxies with {\tt apec+apec} models is similar (0.7 vs 0.8 keV) (to be compared with the typical temperature of hot gas in normal spirals of $\sim 0.3$ keV, \citet{Strickland2004,Mineo2012b}). 
Since the temperatures of the two galaxies
are almost the same, it is tempting to conclude that the physical mechanism which generate the X-ray emission is the same: heating of cold ISM. The ICM ambient temperature 
is higher in ESO137-001 (6keV) than in JW100 (3.5keV), while both the gas density and the thermal pressure are higher for JW100: $3.2 \times 10^{-3} \rm \, cm^{-3}$ vs $1-1.4 \times 10^{-3} \rm \, cm^{-3}$, and $3 \times 10^{-11} \rm \, erg \, cm^{-3}$ vs $1.8 \times 10^{-11} \rm \, erg \, cm^{-3}$.
The ram pressure at the galaxy location for ESO137-001 is $4.4\times 10^{-11}\times (v_{gal}/1500)^2 \rm \, dyn \, cm^{-2}$ and for JW100 is  $1.3\times 10^{-10}\times (v_{gal}/1500)^2 \rm \, dyn \, cm^{-2}$, thus  $ \geq 1.9 \times 10^{-10} \rm \, dyn \, cm^{-2}$ (see Appendix) given that the galaxy peculiar velocity observed with MUSE ($1807 \rm \, km \, s^{-1}$) significantly underestimates its total speed. 

\begin{table*}
\caption{{\bf Results of the X-ray spectral analysis in the control region.}}
\begin{tabular}{ccccc}

\hline
Obs ID. & Exposure time (s) & Bkg exp. time (s) & Total counts (cnts) & Net rate (cnts/s) \\
\hline
16136&1.047e+05&6.725e+05&29992&0.266 (92.7 $\%$ total)\\
3192&2.368e+04&4.615e+05&8445&0.337 (94.4 $\%$ total)\\
\hline
\multicolumn{2}{c}{Model}&\multicolumn{2}{c}{Parameters}&$\chi^2$, DOF, $\chi^2_{R}$\\
\hline
\multicolumn{2}{c}{{\tt phabs*apec}}&\multicolumn{2}{c}{kT=3.50$\pm$0.10 keV, Z=0.36$\pm$0.04}&499.02, 456, 1.0943\\
\hline
\end{tabular}
\end{table*}

\begin{table*}
\caption{{\bf Results of the X-ray spectral analysis in the galactic region.}} 
\begin{tabular}{ccccc}

\hline
Obs ID. & Exposure time (s) & Bkg exp. time (s) & Total counts (cnts) & Net rate (cnts/s) \\
\hline
16136&1.047e+05&6.725e+05&2502&0.0228(95.6 $\%$ total)\\
3192&2.368e+04&4.615e+05&632&0.0260 (97.4 $\%$ total)\\
\hline
\multicolumn{2}{c}{Model}&\multicolumn{2}{c}{Parameters}&$\chi^2$, DOF, $\chi^2_{R}$\\
\hline
\multicolumn{2}{c}{{\tt phabs*apec}}&\multicolumn{2}{c}{kT=1.99$\pm$0.14 keV, Z=0.11$\pm$0.05}&173.43, 96, 1.8066\\
&&&&\\
\multicolumn{2}{c}{{\tt phabs*(apec+apec)}}&\multicolumn{2}{c}{kT=0.82$^{+0.14}_{-0.05}$ keV, (Z=1.00 fixed),}&93.58, 95, 0.9851\\
&&\multicolumn{2}{c}{L$_{0.5-2.0}$=1.99e41 erg/s, L$_{0.5-10.0}$=2.08e41 erg/s  }&\\
&&\multicolumn{2}{c}{L$_{0.3-10.0}$=2.21e41 erg/s }&\\
&&&&\\
\multicolumn{2}{c}{{\tt phabs*(apec+cemekl)}}&\multicolumn{2}{c}{(kT$_{max}$= 3.49keV fixed), (Z=1.00 fixed), $\alpha$=0.88$^{+0.31}_{-0.32}$}&93.84, 95, 0.9878\\
&&\multicolumn{2}{c}{L$_{0.5-2.0}$=3.40e41 erg/s, L$_{0.5-10.0}$=4.54e41 erg/s  }&\\
&&\multicolumn{2}{c}{L$_{0.3-10.0}$=5.00e41 erg/s }&\\
&&&&\\
\multicolumn{2}{c}{$"$}&\multicolumn{2}{c}{kT$_{max}$=1.20$^{+0.51}_{-0.26}$ keV, (Z=1.00 fixed), $\alpha$=2.07$^{+3.32}_{-0.98}$}&87.36, 94, 0.9293\\
&&\multicolumn{2}{c}{L$_{0.5-2.0}$=2.31e41 erg/s, L$_{0.5-10.0}$=2.47e41 erg/s  }&\\
&&\multicolumn{2}{c}{L$_{0.3-10.0}$=2.68e41 erg/s }&\\

\hline
\end{tabular}
\end{table*}


%
%
%
%
%
%
%
%
%
%
%
%

\begin{table*}
\caption{Radio properties at 1.4 GHz. Luminosities are k-corrected. The last column lists the SFR values derived if all the radio continuum luminosity were due to ongoing star formation.}
\begin{tabular}{l|ccc}
\hline
Region  & Flux [10$^{-29}$ W/Hz/m$^2$]&Luminosity [10$^{22}$ W/Hz]&
SFR($M_{\odot} \, yr^{-1}$) \\
\hline
Total (w/out AGN)&3.26$\pm$0.13 (2.36$\pm$0.13)&
2.34$\pm$0.09 (1.69$\pm$0.09)& 7.6(5.5) \\
AGN&0.90$\pm$0.03&
0.65$\pm$0.02& -- \\
Disk (w/out AGN)&2.46$\pm$0.09 (1.56$\pm$0.09)&
1.77$\pm$0.07 (1.12$\pm$0.08)& 5.7(3.6)  \\
Tail&0.8$\pm$0.16&
0.57$\pm$0.12& 1.8 \\
Not compressed (w/out AGN)&1.89$\pm$0.08 (1.59$\pm$0.13)&
1.36$\pm$0.07 (1.14$\pm$0.09)& 4.4(3.7) \\
\hline
\end{tabular}
\vspace{.5cm}
{\newline Note: The regions listed here are the whole radio-emitting regions with a surface brightness above 3 $\sigma$ reported in Fig. 7 (Total), the AGN region identified in the high resolution images \citep[][]{Gitti2013,Ignesti2017} (AGN), the radio-emitting region within the stellar disk as defined by the black contours in Fig.~7 (Disk) and outside the stellar disk (Tail), and the total having excluded the compressed region identified using the $\rm H\alpha$ contours
(Not compressed).}
\end{table*}

\begin{figure*}
\centerline{\includegraphics[width=3.5in,angle=-90]{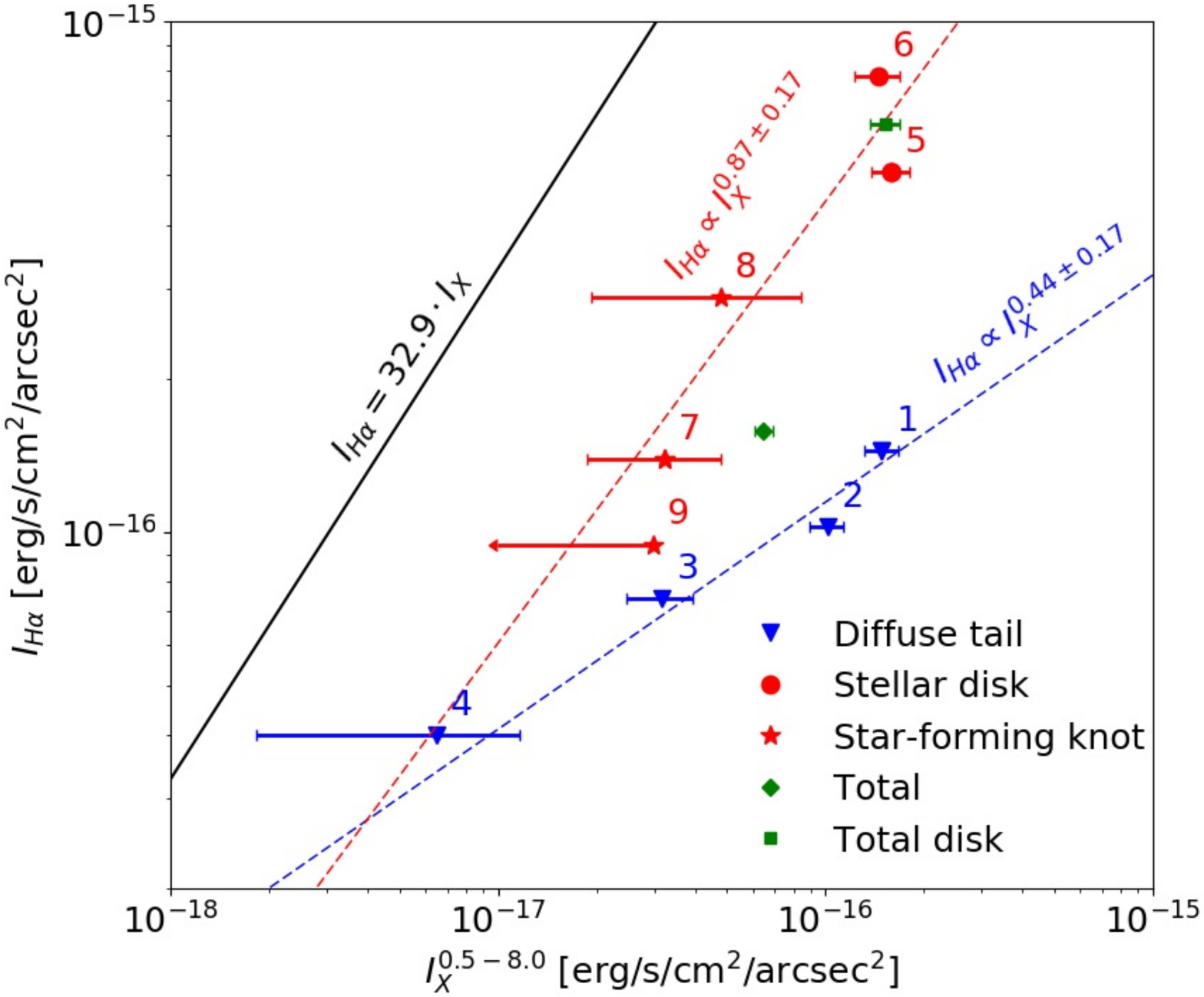}}
\caption{$\rm H\alpha$ vs 0.5-8keV X-ray surface brightness
    in the different regions identified with yellow poligons in
    Fig.~2. The blue triangles are the four regions of diffuse emission with
    LINER-like {\bf [OI]/$\rm H\alpha$} ratios in the
 right panel of Fig.~2. In red, those regions where the [OI]/$\rm H\alpha$ ratio (as well as [SII]/$\rm H\alpha$) indicates star formation: the star-forming regions within
 the disk (red circles)
  and the knots in the tail (red stars). In green, the total and the disk values. The blue dashed line is the fit to the blue diffuse emission points. The solid black line is the expected relation combining the $L_X=SFR$ relation from Mineo et al. (2014) and the $L_{\rm H\alpha}=SFR$ relation from Kennicutt et al. (1998).}
\end{figure*}

\section{X-ray point sources: AGN and ULX candidate }
\label{pointsrc}

\subsection{AGN}
We extract the X-ray AGN spectrum using a background that includes the cluster contribution at the position of the galaxy.\footnote{From the inset in Figure~4
we see that, even using a small region for the extraction of the counts, we expect a residual contamination from the galaxy emission itself, which however is not significant in counts.}
We compute the hardness ratio (HR) defined as the ratio $ \frac{H - S}{H + S} $,
where H and S are the net counts of the source in the hard
(2 - 7 keV) and in the soft (0.5 - 2 keV) band. We find HR = -0.3/-0.10 in the two observations respectively, 
to be compared with the mean ratio for Type 2 AGN HR= -0.03 $\pm$ 0.46 from 
\citet{Marchesi2016}. 
This source is therefore fully consistent with being a low luminosity, absorbed AGN, as confirmed also by the shape of the spectrum which shows two peaks.  We fit the spectrum with a simple combination of a power-law model, plus an absorbed power-law model, fixing the two slopes to be the same, with absorption fixed to the Galactic value of $3.8 \times 10^{20} \rm \, cm^{-2}$ \citep{Kalberla2005}  ({\ttfamily phabs*(pow+phabs+pow)} in xspec).   This shape is a simple representation of a direct plus a reflected component as in an obscured AGN. The resulting slope is $\Gamma = 2.5$ with a very large uncertainty.\footnote{The results do not change significantly by fixing the slope to $\Gamma = 1.7$.} The spectral shape, with a slope steeper than the average slope for Sy2s, is nevertheless very similar to that of the brighter (L$_{X} \sim 10^{43}$ erg/s) Sy2 AXJ2254+1146 \citep{Dellaceca2000}, although the statistics here is not enough to detect a possibile iron line as expected in Seyfert2. The absorbing column is of the order of $3 \times 10^{23} \rm \, cm^{-2}$ and the direct component is about 1\% of the reflected one. The total X-ray luminosity of the AGN is L$_X (0.5-10) = 2.4 \times 10^{41} $ erg/s.


The intensity variability in the 10 years between the two observations is smaller than 10\%, comparable with the statistical uncertainty on the count rate.

\subsection{ULX candidate}
The second, southern X-ray point source (see Table~\ref{tab:pointsrc}) is detected in the X-ray tail, just outside the stellar body of the galaxy. 
Fitting the spectrum with a powerlaw ({\ttfamily phabs*pow} in xspec), 
using again the Galactic absorption value of $3.8 \times 10^{20} \rm cm^{-2}$ \citep{Kalberla2005}, 
we get a
slope $\Gamma = 1.69[1.44-1.96]$ with a $\chi^2_{\nu}$/dof = 1.4/8. 
The corresponding unabsorbed flux (0.5-2 keV) is $3.6 \times 10^{-15} \rm \, erg \, s^{-1} \, cm^{-2}$. 
The flux in the two observations is consistent with no variability. 

The inset in Fig.~4
shows the three X-ray colors image of the galaxy. The two point sources are much harder than the diffuse emission. The AGN is embedded in the galactic emission, while the southern source appears at the edge of the diffuse emission.

We first investigate the possibility that the source is not related to the galaxy. 
The total number of expected contaminants, from the resolved X-ray background LogN-LogS \citep{Moretti2003} at the flux of the source is only 0.027 in an area of 
40$^{\prime\prime} \times 30^{\prime\prime}$, corresponding to the region occupied by the H$\alpha$ emission of the galaxy.
It is therefore unlikely that this is a background source. 
Morever, 
there is no optical counterpart visible in the 
MUSE white-light image, i.e. collapsing all the MUSE wavelength range, down to a magnitude of V $\leq$ 24. Therefore we can exclude the bulk
of AGN counterparts that should have visual magnitudes 
between 22 and 24 \citep{Maccacaro1988}.
Furthermore, 
at the spatial location of this X-ray source
there is no emission line detected in the MUSE data (above the noise) that would reveal the presence of 
an X-ray bright background/foreground source.


We also exclude 
foreground Galactic stars as interlopers:  normal, X-ray emitting stars should be brighter than 20 mag in the optical and would be seen by MUSE, while 
accreting Neutron Stars 
(that would have an X-ray over V-band flux ratio
$f_X$/$f_V$ ratio $\geq$ 1000, corresponding to $m_V$ = 30)
are very rare, especially at high galactic latitude (b$^{\rm II} = -38 ^{\circ}$).

We are left, then, with the interesting possibility that the source is associated to JW100 itself. 
At the luminosity distance of the
cluster, the detected flux corresponds to $L_X (0.3-10) = 7.8 \times 10^{40} \rm \, erg \, s^{-1}$. 
Such a high X-ray luminosity in a point-like source located close to star forming regions make the source an Ultra-Luminous X-ray source (ULX), that is expected to be produced by recent strong star formation episodes (see e.g. \citet{Kaaret2017} 
for a review on ULXs).
The ULX would have formed during the star formation enhancement induced by ram pressure. Interestingly, the ULX candidate is at the location of a bright UV knot (see Fig.~7) and close to an $\rm H\alpha$ clump. Thus, our data shows a spatial association between the ULX candidate and a bright stellar clump recently formed.
At a distance D=255 Mpc this is
the farthest ULX candidate found and a very luminous
one, although not enough to enter the class of Hyper Luminous X-ray sources, i.e. the ULXs with luminosity in excess of  $10^{41} \rm \, erg \, s^{-1}$
see e.g. 
\citep{Gao2003,Wolter2004} for Cartwheel N10
\citep{farrell2009} for HLX-1 in ESO243-49).
The X-ray spectrum is consistent with what seen in other ULXs 
with this level of statistics (eg. \citet{Swartz2011,Wolter2004}).


The bulk of ULXs are consistent with being High Mass X-ray Binaries HMXB. 
We can use the relations from \citet{Mineo2012a} 
concerning the number of HMXB sources and the total X-ray luminosity expected as a function of SFR. 
Using ${\rm SFR}=4.0 M_{\odot}
yr^{-1}$ (see \S2), 
the total number of expected bright sources is $\rm N_{HMBX} (> 10^{39}) = 0.49 \times SFR \sim 2$, which is consistent with our observation of a bright ULX.
However, 
the expected total luminosity in HMXB, of which the majority is non resolved in this observation, 
is ${\rm L_{X}^{HMXB}} = 2.6 \times 10^{39} \times SFR  \; {\rm erg/s} = 1.0 \times 10^{40} {\rm erg/s} $. The detected ULX is already almost an order of magnitude more luminous than the total $L_X$ luminosity expected, outside the scatter in the relation ($\sigma = 0.43 dex $). 

In the galaxy ESO137-001 \citet{Sun2010} identified six ULXs 
with $L_{0.3-10keV}$ up to $2.5\times 10^{40} \rm \, erg \, s^{-1}$, thus much fainter than our candidate ULX in JW100.
The SFR of ESO137-001 is also at least an order of magnitude smaller than in JW100, 
and interestingly the total X-ray luminosity in point sources is larger than expected in both galaxies. 

N-body/smoothed-particle hydrodynamics simulations run with {\ttfamily GADGET-2} have shown that both ram pressure and viscous transfer effects are necessary to produce the large number of ULXs seen in the interacting galaxy NGC 2276 which falls in the potential well and ICM of the NGC 2300 group \citep{Wolter2015}.
We could be witnessing a similar effect here, with 
ram pressure
enhancing the efficiency of the star formation process and the luminosity of the resulting binary system.
A similar effect was found also for another extreme environment, that of collisional ring galaxies, in which both the total number of ULX and the number of ULXs per unit star formation rate are observed to be in the upper envelope of the normal galaxy distribution \citep{Wolter2018}. 
All the evidence suggests a flattening of the X-ray luminosity function when local star formation enhancements/bursts occur,
either due to gravitational interactions and/or stripping,
although statistically strong conclusions cannot be drawn due to the relatively small numbers of ULXs observed.


Finally, if indeed we are witnessing a HMXB, we can confirm that the onset of the interaction with the ICM that triggered the SF episode happened not more than a few hundred Myr ago, given the lifetime of the donor star involved in a HMXB.  



\section{Summary}

In this paper we have studied the jellyfish galaxy JW100 that presents a striking extraplanar tail of multi-phase gas due to ram pressure stripping from the ICM of the Abell 2626 cluster. This work is part of an ongoing effort to understand the physical processes that create tails observable at different wavelengths as well as the baryonic cycle in the tails and disks of jellyfish galaxies.

We use the multi-wavelength dataset of the GASP survey that consists of optical integral-field spectroscopy from MUSE, X-ray ACIS-S Chandra data, 1.4GHz observations from VLA, NUV imaging from UVIT on board ASTROSAT, and CO(2-1) ALMA observations. These data offer a detailed and comprehensive view of the ionized gas, X-ray gas, molecular gas, stellar UV light and radio continuum light emitted from the tail and the disk. The spatial resolution of these observations samples a $\sim$1kpc scale, except for the radio continuum observations that have a resolution $\sim 4 \times 3.5$ kpc. 

Our main results can be summarized as follows:

\begin{itemize}
    \item 
The ICM at the clustercentric distance of JW100 has kT=3.5 keV, a metallicity 0.35 solar, and a density $\rho_{ICM} = 5.8 \times 10^{-27} \rm \, g \, cm^{-3}$.
The X-ray emission of the JW100 region can be equally well modeled adding to the ICM component either an absorbed, thermal, single temperature component (kT=0.82 keV) or a multi-phase, multi-temperature model. The galaxy is moving supersonically (Mach number $\sim$ 2) 
but the presence of a bow shock as inferred by an X-ray temperature break remains unconfirmed until more sensitive observations are obtained.

\item
The 50kpc long $\rm H\alpha$ tail presents bright clumps embedded in diffuse emission. The former are giant and supergiant star-forming regions, mostly located in the south part of the tail. All the MUSE optical line ratios in these clumps indicate that the gas is photoionized by young massive stars. The ionization source of most of the diffuse emission, instead, is star formation according to the [SII] diagnostic diagram, but presents an excess of [OI]-LINER-like emission.

\item
Molecular gas is present in the disk (where $\rm H\alpha$ is also
observed) and in large complexes in the tail, mostly in the southern
part of the tail. A detailed analysis of the ALMA CO emission is given
in Moretti et al. (submitted).

\item
The $\rm H\alpha$ clumps, the CO clumps and the NUV emission are all linked to in situ star formation in the tail, that currently mostly takes place in the south part of the tail.
On small scales (1 to a few kpc), we observe regions with $\rm H\alpha$, CO and UV, but also regions with only bright CO emission, or only $\rm H\alpha$ and UV, or only UV.
We interpret this as a star formation sequence, in which star
formation progresses from the molecular cloud phase with no stars
formed yet, to later stages where the molecular gas has been already
dispersed by the stars formed, until only the UV light of young stars
is still visible. This evolutionary sequence corresponds to a spatial
sequence in the stripped tail, going from further away to closer to the disk.

\item
The radio continuum emission of JW100 is mostly non thermal and is synchrotron emission of relativistic electrons. This indicates the presence of magnetic fields in the stripped tails (see Mueller et al. in prep.
for a direct measurement of magnetic field in a jellyfish tail) and is consistent with supernovae forming in the stripped tails.
The observed 1.4 GHz flux is consistent with that expected from the SFR measured from $\rm H\alpha$ for standard IMF assumptions, though a contribution from other sources such as stripping of relativistic electrons cannot be excluded. 
The spatial distribution of the radio emission, however, does not
coincide with the currently star forming regions: the former is mostly
in the northern part of the tail, and in the south part the radio
avoids the $\rm H\alpha$/CO clumps. This could be due to 
the lifetime of the electrons (longer than the $\rm H\alpha$
timescale), during which they can travel to large distances, and /or
to supernova explosions not having occurred yet in the youngest
star-forming regions.

\item
In contrast with the star formation tracers, the extraplanar X-ray emission is mostly in the northern part of the tail. We find that this X-ray emission cannot be explained 
by star formation (high-mass X-ray binaries and ISM ionized by supernovae and massive stars), because the SFR derived in this case would be between 4 and 30 times higher than the SFR derived from $\rm H\alpha$. The X-ray luminosity observed is similar to the one expected for the hot X-ray halo of a galaxy as massive as JW100, but a simple model rules out the stripped hot halo as the origin of the X-ray tail based on timescale arguments. We conclude that a significant fraction of the X-ray emission of JW100 must arise from heating of its stripped ISM, either due to mixing of the ISM and the ICM, to thermal conduction from the ICM or shock heating.

\item
We find a striking, double correlation between the $\rm H\alpha$ surface brightness and the X-ray surface brightness. The correlation is shallower in regions of diffuse, [OI]-LINER-like emission ({\bf $I_{\rm H\alpha} \propto I_X^{0.44}$}), and steeper in star-forming regions of the disk and tail ({\bf $I_{\rm H\alpha} \propto I_X^{0.87}$}). Even in star-forming regions, the X-ray brightness exceeds significantly the one expected from $\rm H\alpha$ assuming the standard calibrations between SFR and $\rm H\alpha$/UV. 

This result
corroborates the scenario in which the stripped ISM is heated
due to the interaction with the ICM (either mixing, thermal conduction or shocks). This heating could be responsible for: a) much of the X-ray emission; b) the [OI]-excess observed in the diffuse gas of the tail; c) the lack of star formation where such heating is more efficient (i.e. the northern part of the tail). Where the heating is less efficient (in the southern part of the tail, with little or no X-ray emission), star formation occurs in giant clumps with molecular gas. 

     \item
The southern point source is 
most likely a very bright thus rare ULX, 
with a luminosity ($L_X (0.3-10) = 7.8 \times 10^{40} \rm \, erg \, s^{-1}$)
that places it at the bright end of the ULX luminosity function.
Since ram pressure in a dense fluid (as well as gravitational interactions) can enhance the production of bright X-ray sources we deem this a valid explanation for the data in hand. As HMXB (in which the donor star has a short lifetime),
it would have formed during the SF episode triggered at that location by the ram pressure exerted by the ICM a few hundred Myr before observations, consistently with all other evidence for ram pressure stripping in JW100.
     
     \end{itemize}

Multi-wavelength studies of jellyfish galaxies, such as the one we
have presented here for JW100 and those in the literature for
ESO137-001, are powerful probes for a variety of physical processes,
including star formation under extreme environmental conditions
and the interplay between ISM and intergalactic medium. 
Hence, jellyfish galaxies can be a laboratory of physics
for circumgalactic medium (thus galaxy evolution) studies in general.
Fundamental open questions remain unanswered by our results, including how
gas can cool and form new stars in the ICM-embedded tails, and the
physical mechanism of heating of the stripped ISM (how important is mixing vs thermal
conduction vs shocks). Multi-wavelength studies for a larger number of
jellyfish galaxies, for different local ICM conditions, coupled with
hydrodynamical simulations, offer the prospect of significant
advancement in these fields.








\section*{Acknowledgements}
We thank Emiliano Munari for providing us with the CLUMPS code in
advance of publication, Luca Zampieri and Fabrizio Nicastro for useful discussions
regarding the ULX. We are grateful to the anonymous referee who helped us strengthen and clarify the paper. Based on observations collected at the European Organization for Astronomical Research in the Southern Hemisphere 
under ESO programme 196.B-0578. This paper makes use of the following ALMA data: ADS/JAO.ALMA\#2017.1.00496.S. ALMA is a partnership of ESO (representing its member states), NSF (USA) and NINS (Japan), together with NRC (Canada) and NSC and ASIAA (Taiwan), in cooperation with the Republic of Chile. The Joint ALMA Observatory is operated by ESO, AUI/NRAO and NAOJ. This publication uses the data from the AstroSat mission of the Indian Space Research Organisation (ISRO), archived at the Indian Space Science Data Centre (ISSDC). This project has received funding from the European Research Council (ERC) under the European Union's Horizon 2020 research and innovation programme (grant agreement No. 833824). We acknowledge financial support from
PRIN-SKA 2017 (PI L. Hunt). We acknowledge financial contribution from the agreement ASI-INAF n.2017-14-H.0 (PI Moretti), from the INAF main-stream funding programme (PI Vulcani). ER acknowledges the support of STFC, through the University of Hull's Consolidated Grant ST/R000840/1. AW acknowledges financial contribution from the agreement ASI-INAF n.2017-14-H.0. B.V. and M.G. acknowledge the Italian PRIN-MIUR 2017 (PI Cimatti). This work made 
use of the KUBEVIZ software which is publicly available at  
http://www.mpe.mpg.de/$\sim$dwilman/kubeviz/.





\bibliography{gasp_all} 
\bibliographystyle{aasjournal}



\appendix

\section{A model for the stripping of the hot halo }

We can test whether the observed X-ray tail can be due to a hot stripped halo with a simple model. According to \cite{Mccarthy2008} 
the ICM ram pressure strips
the halo gas (assumed spherically distributed) at a projected radius $R$ if

$$ P_{\rm ram} = \rho_{\rm ICM} v_{\rm ICM}^2 > g_{\rm max}(R)
\sigma_{\rm halo}(R)$$

where $\rho_{ICM} = 5.8\times 10^{-27}$ g cm$^{-3}$ is the local ICM
density, $v_{\text ICM}$ is the relative velocity between JW100 and
the ICM, here assumed to lie in the range 1800 (the observed l.o.s velocity) and 2500 km s$^{-1}$ (assumed as fiducial maximum velocity). 
The quantities on the right hand side of the equation are the
maximum gravitational acceleration component parallel to $v_{\text ICM}$ 
at the projected radius $R$ and $\sigma_{\text halo}(R)$ is
the surface density (g cm$^{-2}$) of the hot halo gas.

In order to estimate $g_{max}(R)$ and the distribution of the halo gas
we need to build a mass model for the galaxy. Our fiducial model
includes two stellar components (a stellar bulge and a stellar disk)
and a dark matter halo. The bulge, approximated with a Hernquist
profile \citep{Hernquist1990},
has mass $M_b = 3\times 10^{10}$
M$_\odot$, and half-mass radius $r_{1/2}=1.2$ kpc. The disk has a
very flattened \citep{Miyamoto1975}
distribution
with mass $M_d = 2.5\times 10^{11}$ M$_\odot$ and scale parameters 
$a=6$ kpc and $b=0.5$ kpc (see also \cite{Binney1987}).
The total stellar mass is thus $M_* = 2.8 \times 10^{11}$ M$_\odot$.

Finally, the dark halo assumes a \cite{Navarro1996}
shape,
with total mass $M_{DM}=1.2\times 10^{13}$ M$_\odot$ and a
concentration $c=9.8$.

The hot gas is initially set in hydrostatic equilibrium in the total
potential, assuming a constant temperature $T_{\rm halo} = 5.8\times
10^6$ K ($kT_{\rm halo} = 0.5$ keV), chosen to agree with the mean halo
temperature of NGC 1961 \citep{Anderson2016}.
The central halo gas density is also chosen for the gas density to
agree with the NGC1961 profile \citep{Anderson2016}. The 3D halo
density profile is then integrated to get the surface density
$\sigma_{\text halo}(R)$. 

With this simple model we can use the equation above to calculate the
projected radius $R$ beyond which the stripping is
effective. Figure~11 shows that for $R\ge 380$ pc the ram pressure is larger than the
restoring force per cm$^{-2}$. The time scale to significantly alter the
distribution of the halo gas is $\tau_{\text strip} = 
R/v_{\text ICM} \sim 0.5 (R/{\rm kpc}) (v/{\rm 2000 \,km/s})^{-1}$ Myr. Therefore 
the halo gas is advected by $\sim 30$ kpc (the size of the tail) in $\sim 15$ Myr.
From these numbers we expect that the ram pressure disrupted and removed
most or all the hot halo quickly, well before the galaxy reached the current location,
even taking into account the lower ram pressure experienced by JW100
in the past, while traveling through lower density ICM regions.

In order to gauge the uncertainty in the mass model, we also considered a similar
model as above, replacing the Miyamoto-Nagai disk with a thin Kuzmin disk,
of total mass $M_d^{\rm Kuz} = 2\times 10^{11}$ M$_\odot$ and scale parameter
$a=2.5$ kpc (see \cite{Binney1987}). 
In this case the ram pressure
is able to quickly strip the halo gas beyond $R\sim 950$ pc.

In addition, on top of the classical ram pressure stripping investigated above,
viscous or turbulent ablation \citep{Nulsen1986} would help the gas removal process. To conclude, we expect the hot galaxy halo to be removed on very short timescales, making the hot halo origin of the observed X-ray tail unplausible.


\begin{figure}
\centerline{\includegraphics[width=3.5in]{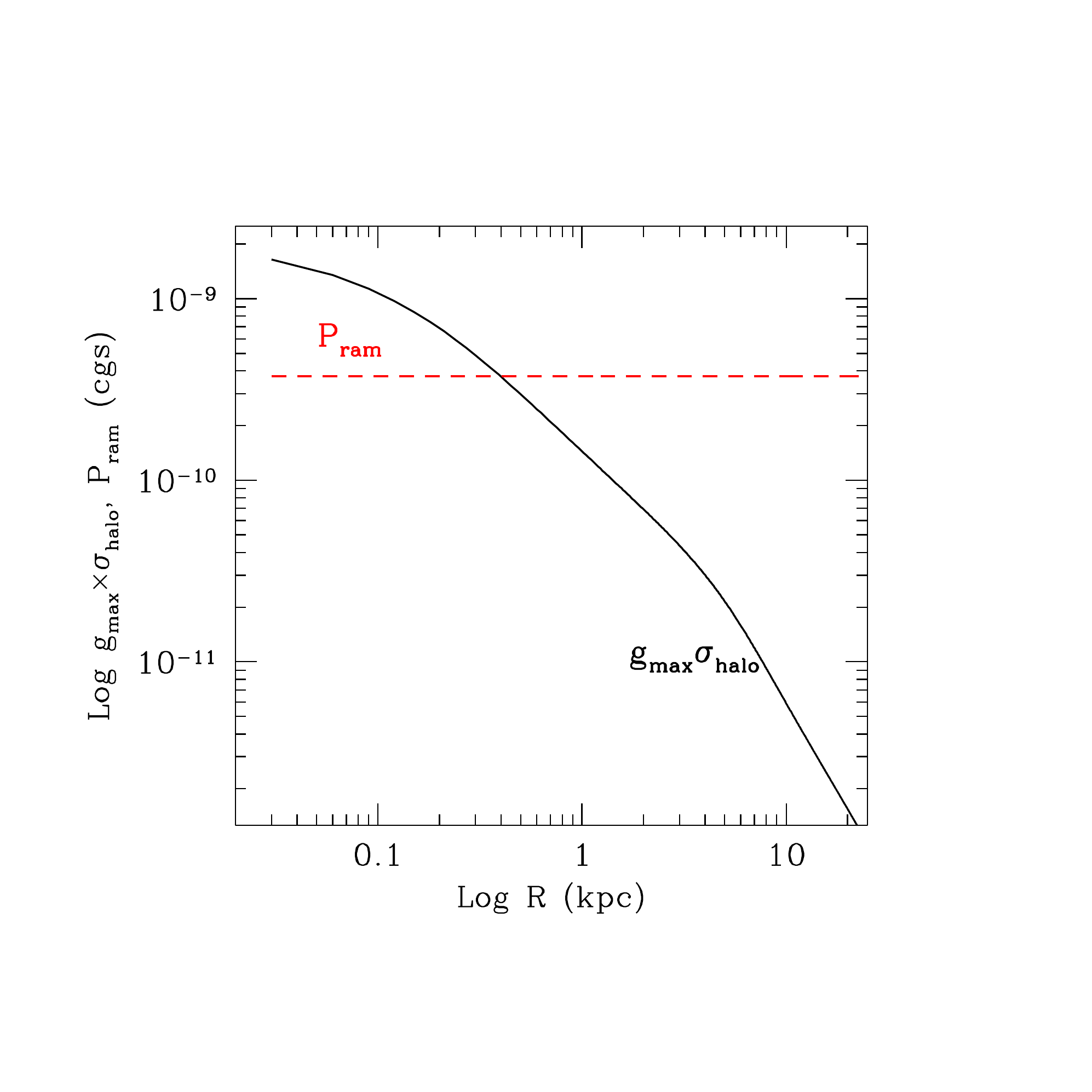}}
\caption{ICM ram pressure (red dashed line) vs. the component along the direction of motion 
of the restoring
force per unit area (solid black line), for the fiducial model with the Miyamoto-Nagai disk.}
\end{figure}


\end{document}